\documentclass[10pt,a4paper]{article}
\usepackage[english]{babel}
\usepackage[utf8]{inputenc}
\usepackage{amsfonts,amsbsy,bm,euscript,mathrsfs}
\usepackage{amssymb,stmaryrd,faktor}
\usepackage[tbtags]{amsmath}
\usepackage{bbm}
\usepackage{graphicx}
\usepackage[title,titletoc]{appendix}
\usepackage[bookmarks=true,colorlinks=true,linkcolor=blue,citecolor=blue,urlcolor=blue,bookmarksnumbered]{hyperref}

\usepackage{dsfont}
\usepackage{collref}
\usepackage{lmodern}
\usepackage{mathrsfs}
\usepackage{mathtools}
\usepackage{bbm}
\usepackage{braket}
\usepackage{slashed}

\usepackage{graphicx}
\usepackage{booktabs}
\usepackage{subfig}
\usepackage{tikz}
\usetikzlibrary{plotmarks,calc,decorations,decorations.pathmorphing}

\numberwithin{equation}{section}

\makeatletter
\renewcommand\section{\@startsection {section}{1}{\z@}
{-3.5ex \@plus -1ex \@minus -.2ex}
{2.3ex \@plus.2ex}
{\normalfont\Large\bfseries}}
\renewcommand\subsection{\@startsection{subsection}{2}{\z@}
{-3.25ex\@plus -1ex \@minus -.2ex}
{1.5ex \@plus.2ex}
{\normalfont\large\bfseries}}
\makeatother

\newcommand{\alg}[1]{\mathfrak{#1}}

\def\z{\zeta}

\def\k{\kappa}

\def\smallL{\small{L}}
\def\smallR{\small{R}}

\newcommand{\bea}{\begin{eqnarray}}
\newcommand{\eea}{\end{eqnarray}}

\newcommand{\cs}{{\rm cs}}
\def\dn{{\rm dn}}

\DeclareMathOperator{\sech}{sech}
\DeclareMathOperator{\arccot}{arccot}

\usepackage{blkarray}

\begin{document}

\thispagestyle{empty}
\begin{flushright}\footnotesize\ttfamily
DMUS-MP-20/02
\end{flushright}
\vspace{2em}

\begin{center}

{\Large\bf \vspace{0.2cm}
{\color{black} \large Boost generator in $AdS_3$ integrable superstrings for general braiding}} 
\vspace{1.5cm}

\textrm{Juan Miguel Nieto Garcia\footnote{\texttt{j.nietogarcia@surrey.ac.uk}}, Alessandro Torrielli \footnote{\texttt{a.torrielli@surrey.ac.uk}} and Leander Wyss\footnote{\texttt{l.wyss@surrey.ac.uk}}}

\vspace{2em}

\vspace{1em}
\begingroup\itshape
Department of Mathematics, University of Surrey, Guildford, GU2 7XH, UK
\par\endgroup

\end{center}

\vspace{2em}

\begin{abstract}\noindent 
In this paper we find a host of boost operators for a very general choice of coproducts in $AdS_3$-inspired scattering theories, focusing on the massless sector, with and without an added trigonometric deformation. We find that the boost coproducts are exact symmetries of the $R$-matrices we construct, besides fulfilling the relations of modified Poincar\'e-type superalgebras. In the process, we discover an ambiguity in determining the boost coproduct which allows us to derive differential constraints on our $R$-matrices. In one particular case of the trigonometric deformation, we find a {non-coassociative} structure which satisfies the axioms of a quasi-Hopf algebra. 

\end{abstract}

\newpage

\overfullrule=0pt
\parskip=2pt
\parindent=12pt
\headheight=0.0in \headsep=0.0in \topmargin=0.0in \oddsidemargin=0in

\vspace{-3cm}
\thispagestyle{empty}
\vspace{-1cm}

\tableofcontents

\setcounter{footnote}{0}

\section{\label{sec:Intro}Introduction}

The existence of integrable structures within the context of the AdS/CFT correspondence \cite{Beisert:2010jr,Arutyunov:2009ga} has allowed us to use a powerful set of techniques that probe both sides of the correspondence beyond perturbation theory. Integrability is made manifest after we map the spectral problem into that of an effective two-dimensional system which is solvable by Bethe ansatz. The associated asymptotic scattering problem, obtained by considering the system in infinite volume, can be studied in the context of Hopf superalgebras, and the problem is reduced to the construction of irreducible representations. However, the one we encounter in this context is very unconventional and is not entirely understood \cite{B1,B2,B3,B4,a,B5,B6,B7}.

The AdS$_3$/CFT$_2$ version of the correspondence has been the focus of some attention in the recent years. This encompasses both the cases of superstrings living on $AdS_3 \times S^3 \times S^3 \times S^1$ and $AdS_3 \times S^3 \times T^4$ backgrounds. The classical integrability of the string sigma-models associated to these backgrounds was shown in \cite{Babichenko:2009dk,Sundin:2012gc} (see also \cite{rev3,Borsato:2016hud}). This has allowed the application of most of the integrability toolbox developed for AdS$_5$/CFT$_4$ for the infinite volume case, while notably the full (massive $+$ massless) Thermodynamic Bethe Ansatz and the Quantum Spectral Curve analysis still need to be developed. Some examples are the construction of the finite-gap equations \cite{OhlssonSax:2011ms} or the construction of the $S$-matrix via the vacuum-preserving subalgebra \cite{Borsato:2012ud,Borsato:2012ss,Borsato:2013qpa,Borsato:2013hoa,Rughoonauth:2012qd,PerLinus,Borsato:2014hja,Borsato:2015mma}, see also \cite{Beccaria:2012kb,Sundin:2013ypa,Bianchi:2013nra,Bianchi:2013nra1,Bianchi:2013nra2}. However, in contrast with excitations in $AdS_5 \times S^5$, excitations in these two backgrounds can be massless. This poses important problems \cite{Sax:2012jv,Borsato:2016xns,Sax:2014mea,Baggio:2017kza}, as massless modes do not follow the same rules as massive excitations \cite{Zamol2,Fendley:1993jh,DiegoBogdanAle}. Prime examples are the modifications needed in the implementation of the Virasoro constraints in the classical algebraic curve \cite{Lloyd:2013wza} or the inaccuracy of the Lüscher corrections to the Asymptotic Bethe Ansatz \cite{Abbott:2012dd,Abbott:2014rca,MI,Abbott:2020jaa}.
 Related work can be found in \cite{Eberhardt:2017fsi,Gaber1,Gaber2,Gaber3,deLeeuw:2020ahe,Gaber4,Gaber5,
GaberdielUltimo,Prin,Prin1,Abbott:2015mla,Per9,Hoare:2018jim,Pittelli:2014ria,Regelskis:2015xxa,Pittelli:2017spf,JuanMiguelAle}.

Since neither the massive nor the massless excitations have a relativistic dispersion relation, a natural question arises regarding the existence of any remnant of Poincar\'e symmetry after gauge-fixing the reparameterisation freedom of the string sigma model. This question was addressed in the context of the scattering problem in $AdS_5 \times S^5$ in \cite{CesarRafa,Charles} and revisited more recently in \cite{BorAle}. We can also pose the same question in the context of the AdS$_3$/CFT$_2$ correspondence, where it has already been studied in \cite{JoakimAle,BorStromTorri,Fontanella:2016opq} for the $AdS_3 \times S^3 \times T^4$ background. In particular, in \cite{FontanellaTorrielli,Fontanella:2019ury} a new variable was found which allows to recast the complete non-relativistic massless $S$-matrix (inclusive of the dressing factor) in a way which not only displays manifest difference form, but it also attains the exact same analytic expression as in the near (left-left and right-right moving) BMN limit \cite{DiegoBogdanAle}. In all cases a suitably modified Poincar\'e symmetry was found and boost operators were identified, although several proposals were made each with different defining features.

Here we will revisit the $AdS_3 \times S^3 \times T^4$ scattering problem from a new perspective and attempt to incorporate the different proposals within one and the same unifying framework. We will also investigate why the difference form achieved by \cite{FontanellaTorrielli,Fontanella:2019ury} seems to be tied to a very specific choice of braiding factor in the coproduct\footnote{We thank Bogdan Stefa\'nski for discussions about this point and for raising this question.}. In doing that we will discover an interesting and unexpected relation between certain ambiguities in the definition of the boost operator and the very structure of the $R$-matrix. Furthermore, we will investigate the existence of the boost in a particular Yang-Baxter deformation of the $AdS_3 \times S^3 \times T^4$ background known as { $\eta$-deformation} \cite{eta1,eta2,eta3,eta4,eta5,eta6,eta7}, see also \cite{Regelskis:2015xxa,eta9,eta10}.

This article is structured as follows. In section 2 we review a single (as sufficient to our purposes) $\alg{su}(1|1)^2$ portion of the vacuum-preserving subalgebra of the full $\alg{su}(1,1|2)^2$ symmetry of the $AdS_3 \times S^3 \times T^4$ background, and the modified Poincar\'e algebra built upon it. In section 3 we will study the different $R$-matrices and boost operators we can construct for different choices of the coproducts of the $\alg{su}(1|1)^2$ generators. In section 4 we perform the same analysis for the $q$-deformed algebra $U_q [\alg{su} (1|1)]$, which is the vacuum-preserving subalgebra of the { $\eta$-deformed} version of the $AdS_3 \times S^3 \times T^4$ background. Here we discover that one of the possible coproducts we can assume generates a non-associative coalgebra, which represents a generalisation of the traditional construction one expects in this case \cite{Drinfeld,z0,z1,z2}. In section 5 we summarise our results and present some concluding comments.

\section{\label{AdS3}Modified Poincaré algebra in AdS$_3$/CFT$_2$}
 
In this section, we present the first algebra we shall be dealing with. This is a non $q$-deformed case, while later on we shall deal with a $q$-deformed algebra. For full detail, we refer to \cite{JoakimAle}, whose conventions we shall follow. We will be concerned with the massless sector of the $AdS_3$ scattering theory, either in the case of $AdS_3 \times S^3 \times T^4$ or $AdS_3 \times S^3 \times S^3 \times S^1$ superstring theory background, since the conclusions we will be making are valid for both. We will consider (one copy of) the $\mathfrak{psu}(1|1)^2$ scattering problem for massless particles, fixing the worldsheet right-movers for definiteness. The dispersion relation of such particles is given by
\begin{equation}
\alg{H}_L = \alg{H}_R = h \sin \frac{p}{2} = \frac{1}{2} E(p) \ ,
\end{equation}
where $E(p)$ is the energy and $p$ the momentum. The other Lie superalgebra generators satisfy the following modified Poincar\'e superalgebra (anti-)commutation relations
\begin{gather}
\{\alg{Q}_A, \alg{S}_A\} \ = \ \alg{H}_A \ ,  \quad [\alg{J}_A, p] \ = \ i \alg{H}_A \ , \quad  [\alg{J}_A, \alg{H}_A] \ = \frac{e^{i p} - e^{-i p}}{2 \mu} \ ,\nonumber\\
[\alg{J}_A, \alg{Q}_A] \ = \frac{i}{2 \sqrt{\mu}} \frac{e^{i \frac{p}{2}} + e^{- i \frac{p}{2}}}{2} \, \alg{Q}_A \ , \qquad [\alg{J}_A, \alg{S}_A] = \frac{i}{2 \sqrt{\mu}} \frac{e^{i \frac{p}{2}} + e^{- i \frac{p}{2}}}{2} \, \alg{S}_A \ , \label{algeq}
\end{gather}
where\footnote{We thank J. Strömwall for pointing out the correct commutation relations. In \cite{JoakimAle} the relations (\ref{algeq}) are extended to the $(L,R),(R,L)$ cases, however the boost-coproduct for mixed handedness $L$ {\it vs.} $R$ does not extend algebraically, but only in the specific representation considered in \cite{JoakimAle}. We shall reserve a detailed analysis of this issue for future work \cite{progress}.} {\footnotesize $A = L,R$} and $\mu = \frac{4}{h^2}$. Apart from the generators we have introduced, we are interested in centrally-extending the algebra in the following way
\begin{equation}
\label{equation}
\{\alg{Q}_{\smallL}, \alg{Q}_{\smallR}\} \ = \ \alg{P} \ , \qquad  \{\alg{S}_{\smallL}, \alg{S}_{\smallR}\} \ = \ \alg{K} \ .
\end{equation}

The fundamental representation we will keep using throughout the paper is the $1|1$-dimensional one, composed of one boson $|\phi\rangle$ and one fermion $|\psi\rangle$. The fermionic charge of such states is accounted for by the { hypercharge operator $\mathcal{B}$}
 \begin{align}
[\mathcal{B},\mathfrak{Q}_L] &= 2 i\mathfrak{Q}_L \ , & [\mathcal{B},\mathfrak{S}_L] &= - 2i \mathfrak{S}_L \ , & [\mathcal{B},\mathfrak{Q}_R] &= -2 i\mathfrak{Q}_R \ , & [\mathcal{B},\mathfrak{S}_R] &= 2i \mathfrak{S}_R \ , \\
& & \mathcal{B}|\phi\rangle &= i|\phi\rangle \ , & \mathcal{B}|\psi\rangle &= -i|\psi\rangle  \ . \label{hypercharge}
\end{align}

From these commutation relations we can write a particularly useful representations of the boost operator $\alg{J}_A$
\begin{equation}
	\alg{J}_A= i \alg{H}_A \, \partial_p \ . \label{differentialformboost}
\end{equation}
This representation is well defined when we consider the two copies separately, however the question on how a boost of one handedness\footnote{We remind that we define {\it handedness} as the $L,R$ label of the algebra generators.} acts on a generator of the opposite handedness proves to be a difficult issue we plan to address it in the future. Meanwhile, in this article we will work only with the fundamental representation\footnote{By the results of \cite{Regelskis:2015xxa} one may argue that the two-dimensional modules are in fact sufficient to construct all other modules in the representation theory of centrally-extended $\mathfrak{su}(1|1)^2$.}, defined by the following action of the supercharges on the states
\begin{equation}
	\alg{S}_L \equiv \alg{Q}_R \equiv \sqrt{\alg{H} (p)} \left( \begin{matrix}
	0 & 1 \\ 0 &0
	\end{matrix} \right) \ , \qquad \alg{Q}_L\equiv \alg{S}_R \equiv \sqrt{\alg{H} (p)} \left( \begin{matrix}
	0 & 0 \\ 1 &0
	\end{matrix} \right) \ , \label{Rep}
\end{equation}
where $\alg{H}_L \equiv \alg{H}_R \equiv \alg{H}$. Thus, combining it with the representation of the boost given above, we can write the following relations
\begin{eqnarray}
\label{algeqc}
&&[\alg{J}_A, \alg{Q}_B] \ = \frac{i}{2 \sqrt{\mu}} \frac{e^{i \frac{p}{2}} + e^{- i \frac{p}{2}}}{2} \, \alg{Q}_B \ , \qquad [\alg{J}_A, \alg{S}_B] = \frac{i}{2 \sqrt{\mu}} \frac{e^{i \frac{p}{2}} + e^{- i \frac{p}{2}}}{2} \, \alg{S}_B \ ,\nonumber \\ 
&&[\alg{J}_{\smallL}, \alg{P}] \ = [\alg{J}_{\smallR}, \alg{P}] \ = \ [\alg{J}_{\smallL}, \alg{K}] \ = [\alg{J}_{\smallR}, \alg{K}]=-[\alg{J}_A, \alg{H}_B]= \frac{e^{- i p} - e^{i p}}{2 \mu} \ ,
\end{eqnarray}
where {\footnotesize $A \neq B$}.

Regarding the coproduct of these generators, one possible choice for the ones for the original $\alg{psu} (1,1|2)^2 \supset \alg{su} (1|1)^2$ algebra is given by
\begin{align}
	\Delta \alg{Q}_A &=\alg{Q}_A \otimes e^{ip/4} +e^{-ip/4}\otimes \alg{Q}_A \ , & \Delta \alg{S}_A &=\alg{S}_A \otimes e^{ip/4} +e^{-ip/4}\otimes \alg{S}_A \ , \notag\\
	\Delta \alg{P} &=\alg{P} \otimes e^{ip/2} +e^{-ip/2}\otimes \alg{P} \ , & \Delta \alg{K} &=\alg{K} \otimes e^{ip/2} +e^{-ip/2}\otimes \alg{K} \ , \notag\\
	\Delta \alg{H}_A &=\alg{H}_A \otimes e^{ip/2} +e^{-ip/2}\otimes \alg{H}_A \ , & \Delta p &=p \otimes \mathfrak{1} +\mathfrak{1} \otimes p \ . \label{copr}
\end{align}
Notice that the Cartan generators $\alg{H}_A$ are also compatible with the coproduct $\tilde{\Delta} \alg{H}_A =\alg{H}_A \otimes e^{-ip/2} + e^{ip/2} \otimes \alg{H}_A$ \cite{FontanellaTorrielli} because we are working in the massless limit, although it may not seem to be consistent with the commutation relations. Precisely, in the massless kinematics one can prove that
\begin{align*}
\label{coprH}
\Delta \alg{H}_A |\Phi \rangle= \left( \alg{H}_A \otimes e^{ip/2} + e^{-ip/2} \otimes \alg{H}_A \right)|\Phi \rangle= h \sin \frac{p_1 + p_2}{2} |\Phi \rangle = \alg{H}_A (\Delta p) |\Phi \rangle \ ,  \\
\tilde{\Delta} \alg{H}_A |\Phi \rangle =\left( \alg{H}_A \otimes e^{-ip/2} + e^{ip/2} \otimes \alg{H}_A \right)|\Phi \rangle = h \sin \frac{p_1 + p_2}{2} |\Phi \rangle= \alg{H}_A (\Delta p)|\Phi \rangle \ ,
\end{align*}
for any state $|\Phi\rangle$.

Regarding the modified Poincaré structure, two different coproducts have been studied for the boost operator in \cite{JoakimAle} and \cite{FontanellaTorrielli}, respectively
\begin{align}
	\Delta_{1} \alg{J}_A &=\alg{J}_A \otimes e^{ip/2} + e^{-ip/2} \otimes \alg{J}_A +\frac{e^{i(p_2-p_1)/4}}{2} \left( \alg{Q}_A \otimes \alg{S}_A + \alg{S}_A \otimes \alg{Q}_A \right) \ , \notag \\
	\Delta_{2} \alg{J}_A &=\alg{J}_A \otimes e^{-ip/2} + e^{ip/2} \otimes \alg{J}_A \ .
\end{align}
We can add to these a third coproduct which, to our knowledge, has not been proposed in the literature yet
\begin{equation}
	{\Delta_3} \alg{J}_A= i (\Delta \alg{H}_A) (\Delta \partial_p)+\text{fermionic tail} \ ,
\end{equation}
where $2\Delta \partial_p=\partial_p \otimes 1 + 1\otimes \partial_p$ and $\Delta \alg{H}_A$ was given in (\ref{copr}). If we substitute in $\Delta_1$ and $\Delta_2$ the differential form of $\alg{J}_A$ given in equation~(\ref{differentialformboost}), we can see that, ignoring the fermionic tails for the moment, they produce differential operators which are distinct from one another, and both distinct from $\Delta_3$. Nevertheless, all three differential operators give the same result when applied to { $p$ and $\alg{H}_A$.}

The reason behind the consistency between these three different coproducts is that the difference between them boils down to a term proportional to $\partial_p \otimes 1 - 1\otimes \partial_p$ (if we disregard again the contribution involving fermionic generators), an operator whose kernel is the polynomial ring in $\Delta p$. Notice that both $\Delta \alg{H}_L$ and $\Delta \alg{H}_R$ live inside this polynomial ring because they fulfil $\Delta [\alg{H}_A (p)]=\alg{H}_A (\Delta p)$. Thus, the bosonic generators of the algebra is annihilated by this operator, and the three choices of coproduct for the boost give the same result when applied to any bosonic generator.

In addition, the term $\partial_p \otimes 1 - 1\otimes \partial_p$ these coproducts differ by is also responsible for the presence or absence of a particular tail involving fermionic generators. This combination of supercharges appears in order to make the coproduct of the boost consistent with the fermionic generators of the algebra. One can easily check that
\begin{equation}
 [4(\partial_p \otimes 1 - 1\otimes \partial_p) , \Delta \alg{Q}_A ]=(\cot (p_1/2) -i) \alg{Q}_A \otimes  e^{ip/4} - (\cot (p_2/2) +i) e^{-ip/4} \otimes \alg{Q}_A \ ,
\end{equation}
and similarly for $\alg{S}_A$. This contribution can be absorbed in the coproduct of the boost operator by adding a fermionic tail of the form
\begin{equation}
	\left( \cot \frac{p}{2} -i \right) e^{ip/4} \alg{Q}_A \otimes e^{ip/4} \alg{H}_A^{-1} \alg{S}_A+ e^{-ip/4} \alg{H}_A^{-1} \alg{S}_A \otimes \left( \cot \frac{p}{2} +i \right) e^{-ip/4} \alg{Q}_A \ .
\end{equation}

We have proven that the three possibilities considered above are acceptable and consistent with the algebra, although they are explicitly different. This forces us to find another criterion to completely and unequivocally fix the coproduct of the boost operator. Therefore, instead of proceeding as in \cite{JoakimAle} and \cite{FontanellaTorrielli}, which eventually produces differential equations satisfied by the massless $AdS_3$ $R$-matrix and determines its difference form, we will choose here to characterise the coproduct by further constraining it to satisfy pure quasi-cocommutativity with the $R$-matrix, i.e.
\begin{equation}
	\Delta^{op} \alg{J}_A R=R\Delta \alg{J}_A \ ,
\end{equation}
where $\Delta^{op}$ is the opposite coproduct, obtained by applying a graded permutation of the two spaces\footnote{This was the approach of \cite{BorStromTorri} for the massive case. The massless case represents a rather singular limit of the coproduct studied in \cite{BorStromTorri}, and in this article we have taken the route of working directly in the massless kinematics. It would be interesting to compare our results with a suitable regularisation of the massless limit described in \cite{BorStromTorri}.}. Interestingly, we will see through the rest of this article that usually it seems to be enough to impose the condition
\begin{equation}
	\langle \phi \phi | \Delta^{op} \alg{J}_A R-R\Delta \alg{J}_A |\phi \phi \rangle = \langle \phi \phi | \Delta^{op} \alg{J}_A -\Delta \alg{J}_A |\phi \phi \rangle=0 \ ,
\end{equation}
for the other equations codified into the quasi-cocommutativity condition to hold. The second equality was obtained using the freedom in choosing a global scalar factor of the $R$-matrix to set $\langle \phi \phi | R |\phi \phi \rangle=1$.

\section{\label{Generalbraiding} General undeformed braiding}

In order to better understand the rôle of the boost $\alg{J}_A$ and how to fix their coproducts, we are going to consider a more general set of allowed coproducts, i.e.
\begin{equation}
	\Delta \alg{S}_L= \alg{S}_L \otimes e^{i a p/4} + e^{i c p/4} \otimes \alg{S}_L \ , \qquad \Delta \alg{Q}_L= \alg{Q}_L \otimes e^{i b p/4} + e^{i d p/4} \otimes \alg{Q}_L \ .
\end{equation}
where $a$, $b$, $c$ and $d$ are constants independent of the momentum\footnote{For the purely algebraic purposes of this paper, there is in principle no restriction to allowing these constant to be complex without worrying about the domain of periodicity of the physical momenta (see footnote 3 in \cite{JoakimAle}). We postpone a thorough treatment of the interplay between the results of this paper and the physical region of momenta to future work. We thank Roberto Ruiz and the referee for discussions about this point.}. The expressions for $\Delta \alg{Q}_R$ and $\Delta \alg{S}_R$ should be constructed so the representation~(\ref{Rep}) is consistent. Notice that we can perform the redefinitions $\alg{S}_L \mapsto e^{i\kappa_S p} \alg{S}_L  $, $\alg{Q}_L \mapsto e^{i\kappa_Q p} \alg{Q}_L$ and $\alg{H}_L \mapsto e^{i(\kappa_S + \kappa_Q) p} \alg{H}_L$  (with $\kappa_S+\kappa_Q=0$ if we want to physically associate the eigenvalue of $\alg{H}$ directly with the (left) one-particle energy) to fix two of the four components $a,b,c,d$. This is because, by the homomorphism property of the coproduct, $\Delta(e^{i z p} T) = \Delta(e^{i z p}) \Delta(T) = e^{i z p} \otimes e^{i z p} \, \Delta(T)$, for any constant $z$ and algebra generator $T$. Thus, only the combinations $a-c$ and $b-d$ are physical and can appear in our final results\footnote{ We shall not allow for Hopf-algebraic two-particle twists of the coproduct at this stage, since they do in general change the physics, therefore they would alter the physical considerations we could eventually draw from this analysis.}. Despite so, we are going to keep the general braidings explicitly, as they will prove important for our discussion.

One important clarification which is necessary at the very start of this analysis is that it is restricted to the massless representation, and it will heavily rely on the specific form of the massless dispersion relation. In this respect, although the general form of the coproduct we start with does not distinguish between massive and massless representations, the conclusions we draw will do. As was the case in \cite{JoakimAle}, it is only because of the specific form $\alg{H} \propto (e^{i \frac{p}{2}} - e^{-i \frac{p}{2}})$ that accidental cancellations occur whereby a non-trivial braiding of the energy can still result in a cocommutative coproduct. This is exemplified by the fact that for instance $\Delta\alg{H}=\alg{H}\otimes e^{i\frac{p}{2}}+e^{-i\frac{p}{2}} \otimes \alg{H} =[ \alg{H}(p_1) e^{i\frac{p_2}{2}}+e^{-i\frac{p_1}{2}} \alg{H}(p_2)] \, \mathfrak{1} \otimes \mathfrak{1}$ is symmetric under the exchange $p_1 \leftrightarrow p_2$ only for the massless form of the energy, and not the massive one. It is in this sense that the conclusions of the present study will not be universal, but only valid for the massless representation. We are planning to address in future work the issue of whether it is possible at all, and, if so, how, to extend to massive particles any of these considerations.

We will focus only on the effects of the boost operator on the fundamental representation previously defined, thus the centrally extended superalgebra $\mathbb{C} \ltimes \mathfrak{su}(1|1)^2 \oplus \mathbb{C}^2$ effectively behaves as just $\mathbb{C} \ltimes \mathfrak{su}(1|1)$, and we can drop the handedness label from our generators.

Starting from our new general form of the coproducts and our representation for the supercharges, we need to recalculate the $R$-matrix, and possibly depart from the actual $AdS_3$ scattering theory for a moment. It turns out that we can still assume that the $R$-matrix has the form
\begin{equation}
	R=\left(\begin{matrix}
	1 & 0 & 0 & 0 \\ 0 & r_{11} & r_{12} & 0 \\ 0 & r_{21} & r_{22} & 0 \\ 0 & 0 & 0 & -1
\end{matrix}	  \right) \ .
\end{equation}
If we impose the quasi-cocommutativity condition with respect to $\alg{Q}$ and $\alg{S}$ we obtain six constraints (we get eight equations but two of them are redundant) although we have only four variables. Hence, this system of equations does not have a solution unless the variables involved in the braiding of the coproduct of the supercharges take some particular values. If we look for such cases, we can indeed identify three different families of consistent braidings\footnote{Although each family has two constraints, only the one associated to the combination $(a-c)+(b-d)$ is physical. Notice that the three families have different values of it, making them inequivalent. In contrast, the other constraint, associated to the combination $a+b+c+d$, arises from fixing the form of the energy and not allowing for a rescaling of it.}
\begin{enumerate}
	\item The family {$a+b=c+d=0$}, which contains as special cases the trivial braiding $a,b,c,d=0$ and the braiding most commonly used in the literature, $a=-b=-c=d=-1$. We will call this family \emph{bosonically unbraided}, as we can always get rid of the braiding in the coproduct of the bosonic Cartan element (the energy) via a redefinition of the generator itself.
	\item The family $a+b=2$ and $c+d=-2$, which corresponds to a generalisation of the coproduct studied in the series of articles \cite{JoakimAle,Fontanella:2016opq, FontanellaTorrielli} and presented in section~\ref{AdS3}. We will call this family \emph{bosonically braided}, as the coproduct of the energy is braided.
	\item The family $a+b=-2$ and $c+d=2$, which can be obtained from the second family via the parity transformation $p\rightarrow -p$.
\end{enumerate}

It is important to stress that the existence of just these three families can be understood in a natural way from imposing the coproduct of the energy $\alg{H}=\{ \alg{Q}, \alg{S}\}=\frac{h}{2} \sin \frac{p}{2}$ to be cocommutative \cite{B2}, i.e., invariant under the exchange of the two spaces\footnote{We remind that having a cocommutative coproduct for a central element is a necessary condition for having an $R$-matrix, since $\Delta^{op}(\mbox{central})R = R\, \Delta(\mbox{central})$ implies $R \, \Delta^{op}(\mbox{central}) = R\, \Delta(\mbox{central})$ because of centrality, hence $\Delta^{op}(\mbox{central}) = \Delta(\mbox{central})$ because of the invertibility of $R$.}

The coproduct of this generator is given by
\begin{equation}
	\Delta \alg{H}=\alg{H} \otimes e^{i (a+b) p/4} + e^{i (c+d) p/4} \otimes \alg{H} \ ,  \qquad \Delta^{op} \alg{H}= e^{i (a+b) p/4} \otimes \alg{H} + \alg{H} \otimes e^{i (c+d) p/4} \ ,
\end{equation}
and imposing $\Delta \alg{H}=\Delta^{op} \alg{H}$ gives us only the three possibilities listed above\footnote{From here we can see that, if we hypothetically took the energy to be instead $\alg{H}\propto e^{i \alpha p/4}- e^{i \beta p/4}$ (with all the necessary changes to the algebraic structure that this would entail), the second and third families would become instead $a+b=\pm \frac{\alpha-\beta}{2}$ and $c+d=\mp \frac{\alpha-\beta}{2}$.}. The other two central elements do not contribute with further constraints because they are equal to $\alg{H}$ in this representation.

In the rest of this section we construct and study in detail the coproduct of the boost operator for the second and first family, respectively.

\subsection{The bosonically braided family}

The bosonically braided case has the added feature of having the coproduct of the energy to be functionally consistent with the one of the momentum \cite{JoakimAle}, namely, $\Delta \mathfrak{H} (p) = \mathfrak{H}(\Delta p)$. Before any analysis on the boost operator, we have first to solve the quasi-cocommutativity condition for the $R$-matrix with respect to the fermionic charges. For this family of parameters, we get\footnote{Notice that the $R$-matrix is not an observable, so the unphysical combination of braiding parameter $x-y\propto a+c$ can appear as long as it does not propagate to the Bethe Equations, which is the case.}
\begin{align}
r_{11} &=\frac{e^{-\frac{i}{4} p_1 (x+y)} \sin \frac{p_2}{2} - e^{-\frac{i}{4} p_2 (x+y)} \sin \frac{p_1}{2} }{\sin \frac{p_1+p_2}{2}} \ , \notag \\
r_{12} &= \frac{2 e^{-\frac{i}{8} (p_1-p_2) (x-y)} \sqrt{\sin \frac{p_1}{2} \, \sin \frac{p_2}{2}}}{\sin \frac{p_1+p_2}{2}} \cos \left( \frac{(x+y)(p_1-p_2) -2 (p_1+p_2)}{8} \right) \ , \notag \\
r_{21} &= \frac{2 e^{\frac{i}{8} (p_1-p_2) (x-y)} \sqrt{\sin \frac{p_1}{2} \, \sin \frac{p_2}{2}}}{\sin \frac{p_1+p_2}{2}} \cos \left( \frac{(x+y)(p_1-p_2) +2 (p_1+p_2)}{8} \right) \ , \notag \\
r_{22} &=- \frac{e^{\frac{i}{4} p_1 (x+y)} \sin \frac{p_2}{2} - e^{\frac{i}{4} p_2 (x+y)} \sin \frac{p_1}{2} }{\sin \frac{p_1+p_2}{2}} \ ,
\end{align}
where we have parameterised our braiding as
$$
a=1+x, \qquad c=-1-y
$$
for convenience. However, despite using the quasi-cocommutativity condition to compute this $R$-matrix, it does not fulfil the Yang-Baxter Equation (YBE) for all the values of the parameters. In fact, only the subfamilies given by $x+y=0$ and $x+y=\pm 2$  fulfil the YBE. These subfamilies of $R$-matrices also satisfy braided unitarity, although only the first one satisfies physical unitarity and behaves well under crossing. Interestingly, the condition $x+y=0$ is also the restriction we have to impose on the $R$-matrix for it to be of difference form (up to phase factors). We elaborate on this in appendix~\ref{diffform}.

To compute the coproduct of the boost operator we can start by writing the ansatz
\begin{multline}
	\Delta \alg{J}= A (p_1 , p_2) \, \left(\partial_{p_1} +\partial_{p_2} \right) +B (p_1 , p_2) \, \left(\partial_{p_1} -\partial_{p_2} \right) \\ +C (p_1 , p_2) \, \alg{S} \otimes \alg{Q} + D (p_1 , p_2) \, \alg{Q} \otimes \alg{S}+  F (p_1 , p_2) \mathcal{B} \otimes \alg{1} + G (p_1 , p_2) \alg{1} \otimes \mathcal{B} \ , \label{boostansatz}
\end{multline}
and impose it to be consistent with the algebra, which can be reduced to impose only the commutation relations of the boost with the momentum and the supercharges. The commutation relation with $\mathcal{B}$ is trivial as our ansatz does not include fermion-number altering terms, while the commutation relation with the energy is fulfilled automatically after imposing the commutation relation with the supercharges.

This procedure does not fix all the functions involved. $A$ is completely fixed, but the equations we obtain can only fix three of the five remaining functions, for example $D$, $F$ and $G$ in terms of $B$ and $C$. Furthermore, the $|\phi \phi \rangle$ component of the quasi-cocommutativity condition is not enough to determine all the remaining ones in this particular case, but it can be used to fix $C$ in terms of $B$. The equation we get in this case is
\begin{align}
	&C (p_1 , p_2)-e^{-\frac{i}{4} (2+x-y) (p_1 -p_2)}C (p_2 , p_1) =
	 \frac{e^{\frac{i}{4} [p_2 (x+1) +p_1 (y-1)]}}{4} \left[ -i \cot \frac{p_1}{2} +i \cot \frac{p_2}{2}  \vphantom{\left( i (x+y) \csc \frac{p_1-p_2}{2}\right) } \right. \notag \\
	&\left. + \frac{[B(p_1 , p_2) - B(p_2 , p_1) ]}{h} \left( i (x+y) \csc \frac{p_1-p_2}{2} -  \csc \frac{p_1}{2} \csc \frac{p_2}{2} \right) \right]\ .
\end{align}
The solution to this equation is
\begin{equation}
	C (p_1 , p_2)= \frac{e^{\frac{i}{4} [p_2 (x+1) +p_1 (y-1)]}}{8} \left( \Upsilon -i \cot \frac{p_1}{2} +i \cot \frac{p_2}{2} +\frac{2 B(p_1 , p_2)}{h} \frac{i (x+y)  -  \cot \frac{p_1}{2} - \cot \frac{p_2}{2}}{\sin \frac{p_1-p_2}{2}} \right) \ ,
\end{equation}
where $\Upsilon=\Upsilon (p_1 , p_2 )$ is an arbitrary symmetric function, i.e. $\Upsilon (p_1 , p_2 )=\Upsilon (p_2 , p_1 )$. Imposing the quasi-cocommutativity condition on the remaining states does not fix this freedom either. We can however fix the function $B$ using the $|\phi \psi \rangle$ component of {said} condition, which amounts to restricting $B$ to be an arbitrary antisymmetric function under the exchange of $p_1$ and $p_2$. Nevertheless, when these two restrictions are imposed, the quasi-cocommutativity condition is fulfilled on all the remaining states as well.

The coefficients involved in the coproduct of the boost operator read then {\footnotesize
\begin{align*}
	A (p_1 , p_2)&= \frac{i h}{2} \sin \frac{p_1+p_2}{2}  \ , \\
	B (p_1 , p_2)&= \frac{i h\beta (p_1 , p_2)}{2} \sin \frac{p_1}{2} \sin \frac{p_2}{2} \ , \notag \\
	C (p_1 , p_2)&= \frac{e^{\frac{i}{4} [p_2 (x+1) +p_1 (y-1)]}}{8} \left[ \Upsilon (p_1 , p_2) -i \cot \frac{p_1}{2} +i \cot \frac{p_2}{2} -i\beta (p_1 , p_2) \frac{\cot \frac{p_1}{2} + \cot \frac{p_2}{2} -i (x+y) }{\cot \frac{p_1}{2} + \cot \frac{p_2}{2}} \right] \ , \notag \\
	 D (p_1 , p_2)&= \notag \\
&\frac{e^{\frac{i}{4} [p_2 (1-x) -p_1 (y+1)]}}{8} \left[ 4-\Upsilon (p_1 , p_2) -i \cot \frac{p_1}{2} +i \cot \frac{p_2}{2} -i\beta (p_1 , p_2) \frac{\cot \frac{p_1}{2} + \cot \frac{p_2}{2} +i (x+y)}{\cot \frac{p_1}{2} + \cot \frac{p_2}{2}} \right]\ , \notag \\
	F (p_1 , p_2)&=\frac{i h}{8} \left[ e^{-\frac{i}{2} p_1}[\Upsilon (p_1 , p_2) -2] \sin \frac{p_2}{2} -x \sin \frac{p_1+p_2}{2} +\beta (p_1 , p_2) \, \frac{x e^{\frac{i}{2} p_2} \sin \frac{p_1}{2} -y e^{-\frac{i}{2} p_1} \sin \frac{p_2}{2} }{\cot \frac{p_1}{2} + \cot \frac{p_2}{2}} \right] \ , \notag \\
	G (p_1 , p_2)&=\frac{i h}{8} \left[ e^{\frac{i}{2} p_2}[\Upsilon (p_1 , p_2) -2] \sin \frac{p_1}{2} +y \sin \frac{p_1+p_2}{2} -\beta (p_1 , p_2) \, \frac{x e^{\frac{i}{2} p_2} \sin \frac{p_1}{2} -y e^{-\frac{i}{2} p_1} \sin \frac{p_2}{2} }{\cot \frac{p_1}{2} + \cot \frac{p_2}{2}} \right] \ , \notag
\end{align*}}
where $\beta(p_1 , p_2 ) =-\beta (p_2 , p_1) $. {We should stress that, although the last two functions explicitly depend on non-physical parameters of the braidings, this is not inconsistent as the boost is probing the momentum structure of the coproduct. An argument in favour of that explanation is the fact that they appear in the functions associated with the outer-automorphisms $\alg{B}$, which can distinguish between the fermionic operators $\alg{Q}$ and $\alg{S}$.

Although} these expressions seem rather complex and lengthy, there exists a point in the parameter space of braidings where they heavily simplify. This point corresponds to the case $x=y$, with a constant $\Upsilon (p_1 , p_2)=2$, and with $\beta(p_1 , p_2)=\cot \frac{p_2}{2} -\cot \frac{p_1}{2} $, and the boost takes the form
\begin{multline}
	\Delta \alg{J}= \Delta_{0} \alg{J} + \frac{e^{-\frac{i}{4} p} \otimes e^{\frac{i}{4} p}}{4} \left( e^{-\frac{ix}{4} p} \alg{S} \otimes e^{\frac{ix}{4} p} \alg{Q} + e^{\frac{ix}{4} p} \alg{Q}  \otimes e^{-\frac{ix}{4} p} \alg{S} \right) - \frac{i x}{8} \left( \cos p \, \mathcal{B} \otimes \alg{H} - \alg{H} \otimes \cos p \, \mathcal{B} \right) \ ,
\end{multline}
where $\Delta_{0} \alg{J}_A= \alg{J}_A \otimes \cos \frac{p}{2} + \cos \frac{p}{2} \otimes  \alg{J}_A$. The coproduct of the boost simplifies even more at the particular braiding $x=y=0$, where we can see that it can be expressed as the average of $\Delta_1 \alg{J}$ and $\Delta_2 \alg{J}$ we defined on section~\ref{AdS3}
\begin{equation}
\Delta \alg{J}= \Delta_{0} \alg{J} + \frac{e^{-\frac{i}{4} p} \otimes e^{\frac{i}{4} p}}{4} \left(  \alg{S} \otimes \alg{Q} + \alg{Q} \otimes \alg{S} \right) \ .
\end{equation}

The two functions that we are not able to fix can be considered ambiguities of the coproduct, as they neither affect the algebra nor the quasi-cocommutativity relation. Their existence can be easily explained from our physical inputs. The {\it first ambiguity} arises from the shared structure of the coproduct of the supercharges, where the constraint imposed by $[\Delta \alg{J}, \Delta \alg{Q}]=\Delta [\alg{J},\alg{Q}]$ is equal to the one imposed by $[\Delta \alg{J}, \Delta \alg{S}]=\Delta [\alg{J},\alg{S}]$. This is also the reason why it is not enough to impose the $|\phi \phi\rangle$ component of the quasi-cocommutativity condition to fix the coproduct of the boost. The {\it second ambiguity} originates from the fact that $\Delta \alg{J}$ cannot be fixed only with the data from the algebra and we have to impose the quasi-cocommutativity condition with the $R$-matrix. As this equation involves the difference between $\Delta \alg{J}$ and $\Delta^{op} \alg{J}$, we have the possibility of adding an \emph{op}-invariant term that does not spoil the commutation relations with the algebra. We will see that other choices of coproducts do not present the first ambiguity, but all of them present the second ambiguity.

If we make both ambiguities explicit, the quasi-cocommutativity relation can be written as
\begin{equation}
	(\Delta^{op} \alg{J} +\hat{\beta}^{op}+\hat{\Upsilon}^{op}) R = R (\Delta \alg{J} +\hat{\beta} +\hat{\Upsilon}) \ ,
\end{equation}
where $\hat{\beta}$ accounts for the first ambiguity, given by
\begin{multline}
	\hat{\beta} = \beta (p_1,p_2) \, \left(\partial_{p_1} -\partial_{p_2} \right) -\beta (p_1,p_2) e^{\frac{i}{4} (p_2 - p_1)} \frac{\csc \frac{p_1}{2} \, \csc \frac{p_2}{2}}{4 h} \\
	\times \frac{\cot \frac{p_1}{2} + \cot \frac{p_2}{2} -i (x+y) }{\cot \frac{p_1}{2} + \cot \frac{p_2}{2}} \,\left[ e^{\frac{i}{4} ( x p_2 + y p_1)} \, \alg{S} \otimes \alg{Q} + e^{-\frac{i}{4} ( x p_2 + y p_1)}  \, \alg{Q} \otimes \alg{S} \right] \ ,
\end{multline}
for any function $\beta (p_1,p_2)$ antisymmetric under the exchange of its arguments, and $\hat{\Upsilon}$ accounts for the second ambiguity, given by
\begin{multline}
	\hat{\Upsilon}= \Upsilon (p_1,p_2) e^{\frac{i}{4} (p_2 -p_1)} \, \left[ e^{-\frac{i}{4} (x p_2 +y p_1)} \alg{S} \otimes \alg{Q} - e^{-\frac{i}{4} (x p_2 +y p_1)} \, \alg{Q} \otimes \alg{S} \right] +\\
	+\Upsilon (p_1,p_2)  \frac{i e^{-\frac{i}{2} p_1}}{2} \, \mathcal{B} \otimes \alg{H} + \Upsilon (p_1,p_2) \frac{i e^{\frac{i}{2} p_2}}{2} \,\alg{H} \otimes \mathcal{B} \ ,
\end{multline}
for any function $\Upsilon (p_1,p_2)$ symmetric under the exchange of its arguments.

Although these ambiguities appear to be inconsequential, they actually seem to encode some information about the structure of the $R$-matrix. For the case $x=y=0$ we can see that the operator encoding the first ambiguity satisfies\footnote{Notice that, although $\hat{\beta}$ is a differential operator, $(\hat{\beta} - \hat{\beta}^{op})$ is not due to its symmetry.}
\begin{equation}
	(\partial_{p_1} - \partial_{p_2})R -  ( \hat{\beta} - \hat{\beta}^{op}) R=0 \ ,
\end{equation}
provided we fix $\beta (p_1 , p_2)= i \cot \frac{p_1 -p_2}{2}$. This equation has already appeared in the literature \cite{Fontanella:2016opq} as an accidental relation for a choice of $\Delta \alg{J}$ that was not quasi-cocommutative, but annihilated the $R$-matrix. As $\hat{\beta}$ quasi-cocommutes with the $R$-matrix, this equation can be written in the fashion of an evolution equation for the $R$-matrix
\begin{equation}
	(\partial_{p_1} - \partial_{p_2})R -  [\hat{\beta}, R]=0 \ .
\end{equation}
On the other hand, the operator encoding the second ambiguity fulfils 
\begin{equation}
	(\partial_{p_1} - \partial_{p_2})R= ( \hat{\Upsilon} - \hat{\Upsilon}^{op}) \ , \label{braidedundeformedupsilonRmatrix}
\end{equation}
provided we fix $\Upsilon (p_1,p_2)=\frac{\cos \frac{p_1-p_2}{2}}{4 ih \sin \frac{p_1}{2} \sin \frac{p_2}{2} \sin \frac{p_1+p_2}{2}}$. Notice the lack of $R$-matrix on the right-hand-side. Sadly, we still lack a first-principle explanation for the origin of these relations.

It is important to stress that the operator dependence of the second ambiguity, $\hat{\Upsilon}$, can be obtained without any knowledge of the $R$-matrix other than the latter admits a boson-boson highest weight entry. Although it might seem otherwise, as we require one component of the cocommutativity equation, we can choose it to be one associated to the highest-weight eigenvalue of the $R$-matrix and use the freedom to multiply the $R$-matrix by a scalar factor to make such eigenvalue equal to one\footnote{It would be interesting to study the issue of normalisation in connection with the physical $AdS_3$ $R$-matrix and the relativistic form of the massless dressing factor \cite{Borsato:2016xns}  displayed in \cite{DiegoBogdanAle,FontanellaTorrielli,Fontanella:2019ury}.}. If we are able to find a first-principle method to find this equation and a method to fix the correct value of $\Upsilon (p_1 , p_2)$, it would be possible to instead compute the $R$-matrix from the ambiguity using equation~(\ref{braidedundeformedupsilonRmatrix}). We expect in fact that the evolution equations we have found can be solved in the same fashion as was done in \cite{Fontanella:2016opq} in terms of path-ordering exponentials for a suitable choice of contours, which should reproduce the expressions of the $R$-matrices we started from.

\subsection{The bosonically unbraided family}

The solution to the quasi-cocommutativity condition with respect to the fermionic charges for this family of coproducts is\footnote{Although it might seems like the components $r_{11}$ and $r_{22}$ depend explicitly on non-physical parameters, this only happens because we should not use the usual relation $E(p)=h \sin \frac{p}{2}$ for the energy. Instead, the { $\alg{H}$ eigenvalue} is given by $\hat{E}(p)=h e^{i (a+b)p/4} \sin \frac{p}{2}$ if we allow for a general rescaling of the supercharges (with $a+b=0$ if we want to physically associate the eigenvalue of the generator $\alg{H}$ directly with the (left) one-particle energy). Substituting this function we see that they depend only on a physical combination of the parameters.}
{\begin{align}
r_{11} &=\frac{ -e^{\frac{i}{4} (b+c) p_2} E (p_1) + e^{\frac{i}{4} (b+c) p_1} E (p_2)}{ e^{\frac{i}{4} (a+b) p_2} E (p_1) + e^{\frac{i}{4} (a+b) p_1} E (p_2)}=\frac{ -e^{\frac{i}{4} (b+c) p_2} \sin \frac{p_1}{2} + e^{\frac{i}{4} (b+c) p_1} \sin \frac{p_2}{2}}{ \sin \frac{p_1}{2} +\sin \frac{p_2}{2}} \ , \notag \\
r_{12} &= \frac{2 e^{\frac{i}{8} [(b+d) p_1 + (a + c) p_2]} \cos \left[ \frac{a-c}{8} (p_1-p_2) \right] \sqrt{E (p_1) E (p_2)}}{ e^{\frac{i}{4} (a+b) p_2} E (p_1) + e^{\frac{i}{4} (a+b) p_1} E (p_2)} \notag \\
	&= \frac{2 e^{\frac{i}{8} [(b+d) p_1 + (a + c) p_2]} \cos \left[ \frac{a-c}{8} (p_1-p_2) \right] \sqrt{\sin \frac{p_1}{2} \, \sin \frac{p_2}{2}}}{\sin \frac{p_1}{2} + \sin \frac{p_2}{2}} \ , \notag \\
r_{21} &= \frac{2 e^{\frac{i}{8} [ (a + c) p_1 + (b+d) p_2]} \cos \left[ \frac{a-c}{8} (p_1-p_2) \right] \sqrt{E (p_1) E (p_2)}}{ e^{\frac{i}{4} (a+b) p_2} E (p_1) + e^{\frac{i}{4} (a+b) p_1}  E (p_2)} \notag \\
	&= \frac{2 e^{\frac{i}{8} [ (a + c) p_1 + (b+d) p_2]} \cos \left[ \frac{a-c}{8} (p_1-p_2) \right] \sqrt{\sin \frac{p_1}{2} \, \sin \frac{p_2}{2}}}{ \sin \frac{p_1}{2} + \sin \frac{p_2}{2}} \ , \notag \\
r_{22} &=\frac{ e^{\frac{i}{4} (a+d) p_2} E (p_1) - e^{\frac{i}{4} (a+d) p_1} E (p_2)}{ e^{\frac{i}{4} (a+b) p_2} E (p_1) + e^{\frac{i}{4} (a+b) p_1} E (p_2)}=\frac{ e^{\frac{i}{4} (a+d) p_2} \sin \frac{p_1}{2} - e^{\frac{i}{4} (a+d) p_1} \sin \frac{p_2}{2}}{\sin \frac{p_1}{2} + \sin \frac{p_2}{2}} \ .
\end{align}}
However, only the following three subfamilies fulfil the Yang-Baxter equation and braided unitarity
\begin{itemize}
	\item The subfamily $a=c$ and $b=d$, which contains the trivial braiding.
	\item The subfamily $a=-b=c-2=-d-2$, which contains the braiding most commonly used in the literature, $a=-b=-c=d=-1$.
	\item The subfamily $a=-b=c+2=-d+2$, which is just the parity transformed of the previous subfamily.
\end{itemize}
Notice that in all these three cases $a+b=c+d$. Regarding physical unitarity and crossing, they always hold for the second and third subfamilies but only hold in the first subfamily if $a+b=0$.

To compute the coproduct of the boost operator we can use the same ansatz we proposed in the previous section
\begin{multline}
	\Delta \alg{J}= A (p_1 , p_2) \, \left(\partial_{p_1} +\partial_{p_2} \right) +B (p_1 , p_2) \, \left(\partial_{p_1} -\partial_{p_2} \right) \\ +C (p_1 , p_2) \, \alg{S} \otimes \alg{Q} + D (p_1 , p_2) \, \alg{Q} \otimes \alg{S}+  F (p_1 , p_2) \mathcal{B} \otimes \alg{1} + G (p_1 , p_2) \alg{1} \otimes \mathcal{B} \ ,
\end{multline}
and impose it to be consistent with the commutation relations involving the momentum and the supercharges, together with the $|\phi \phi\rangle$ component of the quasi-cocommutativity condition for $\alg{J}$. 

In contrast to the previous family of coproducts, here we encounter an ambiguity in how to formulate the constraint imposed by the supercharges because the energy and the momentum have different coproducts. The issue boils down to writing the commutation relation of, e.g. $\alg{Q}$ either as\footnote{There exist several natural choices which still retain a certain degree of simplicity, however that we will not consider them here. The structure constant is actually proportional to $\frac{\sin p}{\alg{H}}$, so four of the simplest cases come from interpreting $\sin p$ as coming from either $\sin p$, $\alg{H} \cos \frac{p}{2}$, $\partial_p \alg{H}^2$ or $\sin \frac{p}{2} \partial_p \alg{H}$.}
\begin{equation}
\label{forms}
	[\alg{J}, \alg{Q}] \ = \frac{i h \cos \frac{p}{2}}{2} \, \alg{Q} \qquad \text{ or as } \qquad [\alg{J}, \alg{Q}] \ = 2 i \partial_p \alg{H} \, \alg{Q} \ ,
\end{equation}
which give rise to two different conditions when applied to the coproducts. One might be tempted to think that the action on the energy $[\Delta \alg{J}, \Delta\alg{H}]=\Delta [\alg{J} , \alg{H}]$ can fix the correct form of the coproduct of $\sin p$, but this is impossible as this relation is automatically satisfied whenever the consistency of the coproduct with the action of the boost on the supercharges is imposed.

If we choose the first form, the expressions for the coefficients involved in the coproduct {of the boost read}
{\footnotesize
\begin{align*}
	A (p_1 , p_2)&= \frac{i h}{2} \left( \sin \frac{p_1}{2} + \sin \frac{p_2}{2} \right)  \ , \\
	B (p_1 , p_2)&= i h \cot \frac{p_1-p_2}{2} \cos \frac{p_1-p_2}{2} \left( \cos \frac{p_1}{2} + \cos \frac{p_2}{2} -2 \cos \frac{p_1+p_2}{2}\right)\ , \\
	C (p_1 , p_2)&= \frac{e^{\frac{i}{4} (a p_2 - c p_1) }}{8} \, \frac{\left( 2i\sec \frac{p_1}{2} \sin \frac{3 p_1}{2} +a -c -\Upsilon (p_1 , p_2) \right) \cos \frac{p_2}{2} + \left( 2i\sec \frac{p_2}{2} \sin \frac{3 p_2}{2} +a -c +\Upsilon (p_1 , p_2) \right) \cos \frac{p_1}{2}}{\cos \frac{p_1}{2} -\cos \frac{p_2}{2}} \ , \\
	D (p_1 , p_2)&= \frac{e^{-\frac{i}{4} (a p_2 - c p_1) }}{8} \, \frac{\left( 2i\sec \frac{p_1}{2} \sin \frac{3 p_1}{2} +a -c +\Upsilon (p_1 , p_2) \right) \cos \frac{p_2}{2} + \left( 2i\sec \frac{p_2}{2} \sin \frac{3 p_2}{2} +a -c -\Upsilon (p_1 , p_2) \right) \cos \frac{p_1}{2}}{\cos \frac{p_1}{2} -\cos \frac{p_2}{2}}  \ , \\
	F (p_1 , p_2)&=\frac{i h}{32} \left( \frac{ 4\cos \frac{p_1 +p_2}{2} \left( a \sin \frac{p_1}{2}+c \sin \frac{p_2}{2} \right)  -2(a+c) \cos \frac{p_1}{2} \sin \frac{p_2}{2}- 2 a \sin p_1+ (a-c) \sin p_2}{\cos \frac{p_1}{2} -\cos \frac{p_2}{2}} +2 \Upsilon (p_1 , p_2) \sin \frac{p_2}{2} \right) \ , \\
	G (p_1 , p_2)&= \frac{i h}{32} \left( \frac{2(a+c) \cos \frac{p_1}{2} \sin \frac{p_2}{2} -4\cos \frac{p_1 +p_2}{2} \left( a \sin \frac{p_1}{2}+c \sin \frac{p_2}{2} \right)+  (a-c) \sin p_1+ 2 c \sin p_2}{\cos \frac{p_1}{2} -\cos \frac{p_2}{2}} +2 \Upsilon (p_1 , p_2) \sin \frac{p_1}{2} \right) \ .
\end{align*}}

Although these expressions seem rather complex and lengthy, the $F$ and $G$ coefficient simplify heavily when $a=-c$.

If we choose the second form, the expressions for the coefficients involved in the {coproduct of the boost read}
{\footnotesize
\begin{align*}
	A (p_1 , p_2)&= \frac{i h}{2} \left( \sin \frac{p_1}{2} + \sin \frac{p_2}{2} \right)  \ , \\
	B (p_1 , p_2)&= -i h \cot \frac{p_1-p_2}{4} \cos \frac{p_1-p_2}{4} \cos \frac{p_1+p_2}{4} \ , \\
	C (p_1 , p_2)&= \frac{e^{\frac{i}{4} (a p_2 - c p_1) }}{8} \left[ 2i \cot \frac{p_1-p_2}{4} \left( \csc \frac{p_1}{2} \csc \frac{p_2}{2} -1 \right) +(a-c) \cot \frac{p_1-p_2}{4} \cot \frac{p_1+p_2}{4} +\Upsilon (p_1 , p_2) \right] \ , \\
	D (p_1 , p_2)&= \frac{e^{-\frac{i}{4} (a p_2 - c p_1) }}{8} \left[ 2i \cot \frac{p_1-p_2}{4} \left( \csc \frac{p_1}{2} \csc \frac{p_2}{2} -1 \right) -(a-c) \cot \frac{p_1-p_2}{4} \cot \frac{p_1+p_2}{4} -\Upsilon (p_1 , p_2) \right]  \ , \\
	F (p_1 , p_2)&= \frac{i h}{32} \, \frac{4a \sin \frac{p_1}{2} \cos \frac{p_2}{2} +2 (\Upsilon (p_1 , p_2) +c-a) \cos \frac{p_1}{2} \sin \frac{p_2}{2} +(a+c-\Upsilon (p_1 , p_2)) \sin p_2}{\cos \frac{p_1}{2} -\cos \frac{p_2}{2}} \ , \\
	G (p_1 , p_2)&= \frac{i h}{32} \, \frac{-4c \cos \frac{p_1}{2} \sin \frac{p_2}{2} +2 (c-a-\Upsilon (p_1 , p_2)) \sin \frac{p_1}{2} \cos \frac{p_2}{2} +(\Upsilon (p_1 , p_2)-a-c) \sin p_1}{\cos \frac{p_1}{2} -\cos \frac{p_2}{2}} \ ,
\end{align*}}
which also heavily simplifies when $a=-c$.

Finally, let us address now the question of the freedom appearing in the coproduct of the boost. The freedom which appears here is exclusively the one due to the fact that the quasi-cocommutativity condition fixes $\Delta \alg{J}$ only up to an $op$-invariant operator. Furthermore, as one of the conditions for this ambiguity is to commute with all the elements of the algebra, it is the same for both forms of the algebra (\ref{forms}).

Such freedom is now given by
\begin{multline}
	\hat{\Upsilon}= \Upsilon (p_1,p_2) e^{\frac{i}{4} (d p_1 +a p_2)} \, \alg{S} \otimes \alg{Q} -\Upsilon (p_1,p_2) e^{\frac{i}{4} (c p_1 +b p_2)} \, \alg{Q} \otimes \alg{S}+\\
	+\Upsilon (p_1,p_2)  \frac{i e^{\frac{i}{4} (a+b) p_1}}{2} \, \mathcal{B} \otimes \alg{H} + \Upsilon (p_1,p_2) \frac{i e^{\frac{i}{4} (a+b) p_2}}{2} \,\alg{H} \otimes \mathcal{B} \ .
\end{multline}
This operator also encodes information about the $R$-matrix as in the previous section, however it does so in a different way, fulfilling instead 
\begin{equation}
\label{instead}
	 (\partial_{p_1} - \partial_{p_2})R=  ( \hat{\Upsilon} - \hat{\Upsilon}^{op}) R \ ,
\end{equation}
{ if} $a=-b=c\mp 2=d=\pm 1$ for $\Upsilon = \frac{\pm e^{\frac{i}{4} (p_1 +p_2)} \left( \sin \frac{p_1}{2} + \sin \frac{p_2}{2} \right)}{4h \sin \frac{p_1}{2} \sin \frac{p_2}{2} \sin \frac{p_1+p_2}{4}}$ . Alternatively, one can have $( \hat{\Upsilon} - \hat{\Upsilon}^{op}) R=0$ if $a=-b=c=-d$, which is independent of the value of $\hat{\Upsilon}$.

In the same fashion as above, we can use the quasi-cocommutativity of the ambiguity to rewrite (\ref{instead}) as an evolution equation for the $R$-matrix
\begin{equation}
	(\partial_{p_1} - \partial_{p_2})R -  [\hat{\Upsilon}, R]=0 \ .
\end{equation}

{ Once again, this last condition allows to reconstruct the form of the $R$-matrix knowing the expression of $\hat{\Upsilon}$.}

\section{\label{Deformedbraiding} $q$-deformed algebra and { $\eta$-deformation }}

In this section we are going to study how the boost operator appears in the case of $U_q [\mathfrak{su} (1|1) ]$ algebra, which is related to the $\eta$-deformation of $AdS_3 \times S^3$ string theories. We refer to the Introduction for references on the subject. We will supplement the algebra presented in \cite{eta7}
\begin{eqnarray}
\label{defalgeq}
\{\alg{Q}, \alg{S} \} = [ \alg{H}]_q \ , \quad && \quad \left[ x \right]_q \equiv \frac{q^x - q^{-x}}{q-q^{-1}} \ .
\end{eqnarray}
with the following action for the boost operator {
\begin{eqnarray}
[\alg{J}, \alg{Q} ] = \phi_Q (p) \alg{Q} = \frac{\beta [2 \alg{H}]_q \sin p}{4 [\alg{H}]_q^2} \alg{Q} \ , \quad && \quad [\alg{J}, \alg{S} ] = \phi_S (p) \alg{S} = \frac{\beta [2 \alg{H}]_q \sin p}{4 [\alg{H}]_q^2} \alg{S} \ , \notag \\
\null [\alg{J}, p] \ = \ \alpha [\alg{H}]_q \ , \quad && \quad [\alg{J}, \alg{H}] \ = \frac{q - q^{-1}}{\log q} \beta \sin p \ , \label{defalgeq2}
\end{eqnarray}}
where $\alpha$ and $\beta$ are constants depending on the deformation parameter $q$ and the coupling constant $h$. We elaborate more on these constants and on the relation between this algebra and the kinematics of excitations in $(AdS_3 \times S^3)_\eta$ in appendix~\ref{etakinematics} in order to justify the form of the commutation relations involving the boost operator. The deformation parameter $q$ is related to the usual deformation parameter $\eta$ in the string theory background as $\log q = -\frac{2\eta}{g (1+\eta^2)}$, where $g$ is the string tension (notice that $h \neq g$ in terms of string theory parameters). Furthermore, in the following we are going to identify $\alg{H}=E/2$ as we did in the undeformed case. We shall also point out that the representation we will utilise is simply obtained from the one we have been using so far, modulo the simple replacement $\mathfrak{H}(p) \to [\mathfrak{H}(p)]_q$.

In this section we will study a coproduct characterised by the following braiding of the supercharges
\begin{gather}
	\Delta \alg{S}= \alg{S} \otimes e^{-a\frac{i}{4} p} q^{-b \frac{E}{4}} + e^{a\frac{i}{4} p} q^{b \frac{E}{4}} \otimes \alg{S} \ , \qquad \Delta \alg{Q}= \alg{Q} \otimes e^{-c\frac{i}{4} p} q^{-d \frac{E}{4}} + e^{c\frac{i}{4} p} q^{d \frac{E}{4}} \otimes \alg{Q} \ , \notag \\
	\Delta p= p\otimes \alg{1} +\alg{1} \otimes p \ , \qquad \Delta [\alg{H}]_q= [\alg{H}]_q \otimes e^{-(a+c)\frac{i}{4} p} q^{(b+d) \frac{E}{4}} +e^{(a+c)\frac{i}{4} p} q^{-(b+d) \frac{E}{4}} \otimes [\alg{H}]_q \ ,
\end{gather}
where $a$, $b$, $c$ and $d$ are integer numbers. Instead of looking at the general case, here we will only consider two cases: either $a=b=-c=d=1$, which will give us a trivial braiding for the energy; and $a=b=c=-d=1$, which makes the coproduct of the Cartan element of the algebra consistent with the coproduct of the momentum. The first coproduct and the $R$-matrix associated to it correspond to the massless limit of the ones studied in \cite{eta7} (see also \cite{H1,H2,H3,IRF,Pachol:2015mfa,Hernandez:2020igz,Seibold:2019dvf}) while the second one has not appeared in the literature yet to our knowledge.

\subsection{\label{CoproductBen} Unbraided energy case}

The solution to the quasi-cocommutativity condition with respect to the fermionic charges for the case $a=b=-c=d=1$ reads
\begin{align}
r_{11} &=\frac{e^{\frac{i p_1}{2}} \left[ \frac{E_2}{2} \right]_q -e^{\frac{i p_2}{2}} \left[ \frac{E_1}{2} \right]_q }{ \left[ \frac{E_1 +E_2}{2} \right]_q} \ , \notag \\
r_{12} &= \frac{ \left( e^{\frac{i }{2} (p_1 -p_2)} q^{- \frac{E_1 + E_2}{4}} +e^{-\frac{i}{2} (p_1 -p_2)} q^{ \frac{E_1 + E_2}{4}} \right) \sqrt{\left[ \frac{E_1}{2} \right]_q \left[ \frac{E_2}{2} \right]_q}}{ \left[ \frac{E_1 +E_2}{2} \right]_q} \ , \notag \\
r_{21} &= \frac{ \left( e^{\frac{i }{2} (p_1 -p_2)} q^{\frac{E_1 + E_2}{4}} +e^{-\frac{i}{2} (p_1 -p_2)} q^{ - \frac{E_1 + E_2}{4}} \right) \sqrt{\left[ \frac{E_1}{2} \right]_q \left[ \frac{E_2}{2} \right]_q}}{ \left[ \frac{E_1 +E_2}{2} \right]_q} \ , \notag \\
r_{22} &=\frac{e^{\frac{i p_2}{2}} \left[ \frac{E_1}{2} \right]_q - e^{\frac{i p_1}{2}} \left[ \frac{E_2}{2} \right]_q}{ \left[ \frac{E_1 +E_2}{2} \right]_q} \ .
\end{align}

Notice that this choice of coproduct is compatible with the $q$-analog structure, meaning that $\Delta [\alg{H}]_q=[\Delta \alg{H}]_q$, which implies that the coproduct of the energy is trivial.

To compute the coproduct of the boost operator we can start by writing the ansatz
\begin{multline}
	\Delta \alg{J}= A (p_1 , p_2) \, \left(\partial_{p_1} +\partial_{p_2} \right) +B (p_1 , p_2) \, \left(\partial_{p_1} -\partial_{p_2} \right) \\ +C (p_1 , p_2) e^{i \frac{p_1+p_2}{4}} \, \alg{S} \otimes \alg{Q} + D (p_1 , p_2) e^{-i \frac{p_1+p_2}{4}} \, \alg{Q} \otimes \alg{S}+  F (p_1 , p_2) \mathcal{B} \otimes \mathbb{I} + G (p_1 , p_2) \mathbb{I} \otimes \mathcal{B} \ ,
\end{multline}
and impose it to be consistent with the algebra and the quasi-cocommutativity condition. We have extracted a factor $e^{\pm i \frac{p_1+p_2}{4}}$ from coefficients $C$ and $D$ for convenience. It is important to stress that imposing just the $|\phi \phi \rangle$ component of the quasi-cocommutativity condition (together with consistency with the algebra) is enough to completely fix all the coefficients of this ansatz and make it consistent with the remaining components. Furthermore, following the same reasoning as in the undeformed case, the consistency with the algebra can be reduced to imposing only the commutation relations of the boost with the momentum and the supercharges. Sadly, similarly to the second family of coproducts we studied in the previous section, here we also find the ambiguity arising from the definition of the coproduct of $\sin p$, as it could for instance be interpreted as either $\sin (\Delta p)$, $\Delta \left[ \frac{E}{2} \right]_q \cos (\Delta \frac{p}{2})$, $\sin (\Delta \frac{p}{2}) \Delta \partial_p \Delta \left[ \frac{E}{2} \right]_q$ or $\Delta \partial_p \Delta \left[ \frac{E}{2} \right]_q^2$, if we only consider relatively simple combinations. We have found that all of these four choices gives us a consistent coproduct of the boost. It would be interesting to investigate more the consequences of this choice.

Here we will only collect the expression for the coefficients $A$, $B$ and $C-D$ for any choice of this coproduct, referred to as just $\Delta \sin p$,
\begin{align}
	A(p_1 , p_2) &=\frac{\alpha}{2} \Delta [E]_q=\frac{\alpha}{2} [E_1+E_2]_q  \\
	B(p_1 , p_2) &=  \frac{\alpha}{2} \, \frac{2 (\Delta \sin p)- [E_1+E_2]_q \left( \frac{\sin p_1}{[E_1]_q} -\frac{\sin p_2}{[E_2]_q} \right)}{\frac{\sin p_1}{[E_1]_q} -\frac{\sin p_2}{[E_2]_q}} \\
	\frac{C-D}{2} &= q^{\frac{E_1 -E_2}{4}} \, \frac{i B(p_1 , p_2) +\Upsilon (p_1 , p_2)}{\left[ \frac{E_1 + E_2}{2} \right]_q} \ .
\end{align}
The reason for choosing these three coefficients are that $A$ is the only coefficient completely independent of the choice of the coproduct of $\sin p$, $C-D$ shows how the ambiguity $\hat{\Upsilon}$ is still present in this case, and $B$ to exhibit how $\Delta \sin p$ enters in the computations.

The ambiguity we have { encountered } here has the same origin as the second kind of ambiguities that appeared in the undeformed case. In this case it is given by
\begin{equation}
	\hat{\Upsilon} = \frac{ q^{\frac{E_1 -E_2}{4}} \Upsilon (p_1 , p_2)}{\left[ \frac{E_1 + E_2}{2} \right]_q} \left( e^{i \frac{p_1+p_2}{4}} \,  \alg{S} \otimes \alg{Q} + e^{-i \frac{p_1+p_2}{4}} \, \alg{Q} \otimes \alg{S} + \frac{q^{\frac{E_1 -E_2}{4}}}{2 i} \mathcal{B} \otimes  \left[ \alg{H} \right]_q + \frac{q^{-\frac{E_1 -E_2}{4}}}{2 i} \left[ \alg{H} \right]_q \otimes \mathcal{B} \right) \ .
\end{equation}
However, we still have not found a relation between this ambiguity and the $R$-matrix.

\subsection{\label{CoproductJuan} Braided energy case}

As in the undeformed case, the $q$-analog of energy is functionally consistent with the momentum, namely, $\Delta [E/2]_q = [E/2]_q(\Delta p)$. However, in contrast with that case, this is the only structure consistent with the momentum due to the non-linearity of it.

The solution to the quasi-cocommutativity condition with respect to the fermionic charges for the case $a=b=c=-d=1$ reads
\begin{align}
r_{11} &=\frac{ q^{-\frac{E_1}{2}} \sin \frac{p_2}{2} -q^{-\frac{E_2}{2}} \sin \frac{p_1}{2}  }{\sin \frac{p_1 +p_2}{2}}\ , \notag \\
r_{12} &= \frac{ \left( e^{\frac{i }{4} (p_1 +p_2)} q^{-\frac{E_1 - E_2}{4}} +e^{-\frac{i }{4} (p_1 +p_2)} q^{\frac{E_1 - E_2}{4}}  \right) \sqrt{\sin \frac{p_1}{2} \sin \frac{p_2}{2}}}{\sin \frac{p_1 +p_2}{2}}   \ , \notag\\
r_{21} &= \frac{ \left( e^{\frac{i }{4} (p_1 +p_2)} q^{\frac{E_1 - E_2}{4}} +e^{-\frac{i }{4} (p_1 +p_2)} q^{-\frac{E_1 - E_2}{4}}  \right) \sqrt{\sin \frac{p_1}{2} \sin \frac{p_2}{2}}}{\sin \frac{p_1 +p_2}{2}}  \ , \notag \\
r_{22} &=\frac{ q^{\frac{E_2}{2}} \sin \frac{p_1}{2} -q^{\frac{E_1}{2}} \sin \frac{p_2}{2} }{\sin \frac{p_1 +p_2}{2}} \ .
\end{align}

To compute the coproduct of the boost operator we can start by writing the ansatz
\begin{multline}
	\Delta \alg{J}= A (p_1 , p_2) \, \left(\partial_{p_1} +\partial_{p_2} \right) +B (p_1 , p_2) \, \left(\partial_{p_1} -\partial_{p_2} \right) \\ +C (p_1 , p_2) q^{-\frac{E_1+E_2}{4}} \, \alg{S} \otimes \alg{Q} + D (p_1 , p_2) q^{\frac{E_1+E_2}{4}} \, \alg{Q} \otimes \alg{S}+  F (p_1 , p_2) \mathcal{B} \otimes \mathbb{I} + G (p_1 , p_2) \mathbb{I} \otimes \mathcal{B} \ ,
\end{multline}
and impose it to be consistent with the algebra and the $|\phi \phi \rangle$ component of the quasi-cocommutativity condition. In this case it is more convenient to extract a factor $q^{\pm \frac{E_1+E_2}{4}}$ from coefficients $C$ and $D$. 

The coefficients involved in the coproduct of the boost read in this case 
{\footnotesize
\begin{align*}
	A(p_1 , p_2 ) &= \frac{\alpha}{2} [E]_q (p_1 +p_2 ) \ , \\
	B(p_1 , p_2 ) &\propto  A(p_1 , p_2 ) \left[ \left( \frac{\phi_Q (p_1)}{\alpha [E_1]_q}+\frac{i}{4} \right) e^{\frac{i}{4} (p_1 + p_2)} \left[\frac{E_1}{2} \right]_q + \left( \frac{\phi_Q (p_2)}{\alpha [E_2]_q}-\frac{i}{4} \right) e^{-\frac{i}{4} (p_1 + p_2)} \left[\frac{E_2}{2} \right]_q  \right] \notag \\
	&-\phi_Q (p_1 + p_2) \left[ e^{\frac{i}{4} (p_1 + p_2)} \left[\frac{E_1}{2} \right]_q + e^{-\frac{i}{4} (p_1 + p_2)} \left[\frac{E_2}{2} \right]_q  \right]=0 \ , \\
	C (p_1 , p_2) &= -\frac{1}{2i} \left( A (p_1 , p_2 )  \log q \frac{\partial_{p_2} E_2}{4} +\frac{F-G}{2} e^{-\frac{i}{4} (p_1 + p_2)} \left[\frac{E_2}{2} \right]_q  \right) \ , \\
	D (p_1 , p_2) &= \frac{1}{2i} \left( A (p_1 , p_2 )  \log q \frac{\partial_{p_1} E_1}{4} -\frac{F-G}{2} e^{\frac{i}{4} (p_1 + p_2)} \left[\frac{E_1}{2} \right]_q  \right) \ , \\
	\frac{F+G}{2} &=- e^{-\frac{i}{4} (p_1 - p_2)} \frac{i A(p_1 , p_2 ) }{2 \left[ \frac{E (p_1 + p_2)}{2} \right]_q} \left( 2-i \cot \frac{p_1}{2} +i \cot \frac{p_2}{2} \right) \ , \\
	\frac{F-G}{2} &= \frac{e^{-\frac{i}{4} (p_1 - p_2)}}{\left[ \frac{E (p_1 + p_2)}{2} \right]_q} \left( A(p_1 , p_2 ) \log q \frac{\partial_{p_1} E_1 -\partial_{p_2} E_2}{4} +\Upsilon (p_1 , p_2) \right) \ .
\end{align*}}

This case is rather special. Imposing the $|\phi \phi \rangle$ component of the quasi-cocommutativity condition to vanish fixes all the functions appearing in our ansatz but the resulting form of the coproduct do not fulfil the remaining components of said condition. Furthermore, the $\hat{\Upsilon}$ operator does not seems to quasi-cocommute with the $R$-matrix. We still do not understand why this happens or why it only happens in this case. However, we suspect that it has to do with the algebra not being {coassociative} (see the next section).

The simplest procedure to see that the algebra is not coassociative is to compute the dependence of $q^{\frac{E}{4}}$ on $\left[ \frac{E}{2} \right]_q$
\begin{equation}
	\left[ \frac{E}{2} \right]_q=\frac{q^{\frac{E}{2}}-q^{-\frac{E}{2}}}{q-q^{-1}} \Longrightarrow q^{\frac{E}{4}}= \sqrt{\frac{\left[ \frac{E}{2} \right]_q (q-q^{-1}) +\sqrt{4+\left[ \frac{E}{2} \right]_q^2 (q-q^{-1})^2}}{2}} \ .
\end{equation}
We can now construct a formal coproduct of $q^{\frac{E}{4}}$ if we expand the previous expression in a Taylor series in $\left[ \frac{E}{2} \right]_q$, compute the coproduct order by order, and sum back the series. Nevertheless, by direct inspection we can conclude that the coproduct of $q^{\frac{E}{4}}$ is not group-like, namely $\Delta q^{\frac{E}{4}} \neq q^{\frac{E}{4}} \otimes q^{\frac{E}{4}}$. As the coproduct of the supercharges involve this factor, this implies that $(\mathfrak{1} \otimes \Delta) \Delta \alg{Q} \neq (\Delta \otimes \mathfrak{1}) \Delta \alg{Q}$ and $(\mathfrak{1} \otimes \Delta) \Delta \alg{S} \neq (\Delta \otimes \mathfrak{1}) \Delta \alg{S}$.

\subsection{Quasi-Hopf algebra and the coassociator}

In the previous section we have verified that a particular choice of coproducts is not coassociative due to $\Delta q^{\frac{E}{4}}$ not being group-like. In this section we show that this coproduct rather satisfies the properties which characterise {\it quasi-Hopf} algebras \cite{Drinfeld,z2,QuasiHopf}\footnote{We denote the identity always as $\mathfrak{1}$ in any context, whether it is the identity map or the identity matrix of any dimension.}. In these structure the coassociativity is replaced by the relation
\begin{equation}
(\mathfrak{1} \otimes \Delta) \Delta (x) = \Phi . \Big[(\Delta \otimes \mathfrak{1}) \Delta(x)\Big] . \Phi^{-1} \ ,
\end{equation}
for all algebra elements $x$, with the {\it coassociator} $\Phi$ being a certain invertible { element } in the triple tensor product space subject to a list of conditions \cite{Drinfeld,QuasiHopf}. In our case, given the hypercharge element $\mathcal{B}$ defined in equation~(\ref{hypercharge}), one can prove that the suitable coassociator is given by 
\begin{eqnarray}
\label{prova}
\Phi = \exp (\mathcal{B} \otimes \omega_{23} + \omega_{12} \otimes \mathcal{B}), \qquad \omega_{jk} = \frac{1}{8 i} \log q \, \Big(E(p_j + p_k) - E(p_j) - E(p_k)\Big) \mathfrak{1} \otimes \mathfrak{1} \ , 
\end{eqnarray}
where the indices in $\omega_{jk}$ refer as usual to the appropriate spaces in the triple tensor product. Notice that one can abstractly write
\begin{equation}
\omega = \Delta(E) - \Delta_{\text{trivial}} (E) \ , 
\end{equation}
which displays how the coassociator is measuring the departure of the energy-coproduct from the trivial coproduct $\Delta_{\text{trivial}}(E) = E \otimes \mathfrak{1} + \mathfrak{1} \otimes E$, which is precisely a measure of the non-group-like nature of $\Delta q^{\frac{E}{4}}$.
The proof of (\ref{prova}) relies on the properties
\begin{eqnarray}
e^{\alpha \mathcal{B}} \, \mathfrak{Q} = \mathfrak{Q} \, e^{\alpha (\mathcal{B}+ 2 i \mathfrak{1})},  \qquad e^{\alpha \mathcal{B}} \, \mathfrak{S} = \mathfrak{S} \, e^{\alpha (\mathcal{B}- 2 i \mathfrak{1})} \ ,
\end{eqnarray}
and on the fact that $\omega_{ij}$ is a central element. Moreover, it is easy to see that the coassociator satisfies the properties listed in \cite{Drinfeld}, and therefore it qualifies the coproduct to constitute a quasi-Hopf algebra. In particular, by noticing that 
\begin{eqnarray}
&&(\Delta \otimes \mathfrak{1}) (\omega_{ij}) = \frac{1}{8i} \log q \, \Big(E(p_i + p_j + p_k) - E(p_i + p_j) - E(p_k)\Big) \mathfrak{1} \otimes \mathfrak{1} \otimes \mathfrak{1} \ , \nonumber \\
 &&(\mathfrak{1} \otimes \Delta) (\omega_{ij}) = \frac{1}{8i} \log q \, \Big(E(p_i + p_j + p_k) - E(p_i) - E(p_j + p_k)\Big) \mathfrak{1} \otimes \mathfrak{1} \otimes \mathfrak{1} \ ,
\end{eqnarray}
one can prove the so called \emph{pentagon relation}
\begin{eqnarray}
\Big[(\mathfrak{1} \otimes \mathfrak{1} \otimes \Delta) \Phi \Big]\, . \, \Big[(\Delta \otimes \mathfrak{1} \otimes \mathfrak{1}) \Phi\Big] = (\mathfrak{1} \otimes \Phi)\,  . \, \Big[(\mathfrak{1} \otimes \Delta \otimes \mathfrak{1}) \Phi\Big] \, . \, (\Phi \otimes \mathfrak{1}) \ .
\end{eqnarray}
Furthermore, the counit $\epsilon$ annihilates all Lie superalgebra elements, hence $\epsilon(p) = 0$ and $\epsilon(E) = 0$ (unlike what happens for group-like elements, {\it e.g.} $\epsilon(\mathfrak{1}) = 1$), whence one gets
\begin{eqnarray}
(\mathfrak{1} \otimes \epsilon \otimes \mathfrak{1}) \Phi = \mathfrak{1} \otimes \mathfrak{1} \ , 
\end{eqnarray}
in addition to the property already in effect 
\begin{eqnarray}
(\mathfrak{1} \otimes \epsilon) \Delta = (\epsilon \otimes \mathfrak{1}) \Delta = \mathfrak{1} \ .
\end{eqnarray}

One can see that the antipode relations for the coassociator work as well. To begin with, we notice that our Lie superalgebra generators, say, $x$, all satisfy the purely Hopf-algebra property
\begin{equation}
\mu (S \otimes \mathfrak{1}) \Delta (x) = \eta \epsilon (x) = 0 = \mu (\mathfrak{1} \otimes S) \Delta (x) \ ,
\end{equation}
where $S$ is the antipode, $\eta$ the unit, $\epsilon$ the counit and $\mu$ the multiplication. This can be applied to the coproduct of the momentum, and one derives
\begin{equation}
S(p) = - p, \qquad S(E) = - E \ ,
\end{equation} 
the last property being obtained by using the fact that $S$ is an (anti)homomorphism, and recalling that $[\frac{E}{2}]_q$ is proportional to $\sin \frac{p}{2}$, hence odd in $p$, together with the relation $[-\frac{E}{2}]_q = - [\frac{E}{2}]_q$.

For our quasi-bialgebra to be a quasi-Hopf algebra we need to prove that the coassociator satisfies the (simplified) antipode-coassociator relations
\begin{equation}
\label{see}
\mu (\mathfrak{1} \otimes \mu) \Big[S \otimes \mathfrak{1} \otimes S\Big] \Phi^{-1} = 1 = \mu (\mathfrak{1} \otimes \mu) \Big[\mathfrak{1} \otimes S \otimes \mathfrak{1}\Big] \Phi,
\end{equation}
where $\mu (\mathfrak{1} \otimes \mu) = \mu (\mu \otimes \mathfrak{1})$ for our associative multiplication (which of the two ways can therefore be chosen as it is most convenient). 
In order to verify (\ref{see}), it is sufficient to notice how the antipodes acting on the coassociator in the fashion displayed above will always only involve one space at a time where the energy is sitting, hence one will always end up, for both the coassociator and its inverse, evaluating at the exponent combinations such as
\begin{equation}
\label{annih}
E(-p_i+p_j) + E(p_i) - E(p_j).
\end{equation}
We then need to remember that the final result of the multiplications will be to force the evaluation to be on one and the same space, hence eventually one needs to set $p_1 = p_2 = p_3$, which annihilates the combination (\ref{annih}), causing all exponents to vanish and completing the proof of (\ref{see}). This shows that the structure we are dealing with in this section is a true quasi-Hopf algebra.

In much the same vein as the Hopf algebra structure can be generalised to the one of a quasi-Hopf algebra, there is also a non-coassociative generalisation of the quasi-triangularity structure and the Yang-Baxter equation. For a quasi-triangular quasi-Hopf algebra, the universal $R$-matrix fulfils the \textit{generalised Yang-Baxter equation}
\begin{equation} \label{generalizedYB}
R_{12} \Phi^{312} R_{13} {\Phi^{132}}^{-1} R_{23} \Phi^{123} = \Phi^{321} R_{23} {\Phi^{231}}^{-1} R_{13} \Phi^{213} R_{12} \ , 
\end{equation}
where $\Phi^{\sigma} = \sum_i \Phi_i^{\sigma^{-1} (1)} \otimes \Phi_i^{\sigma^{-1} (2)} \otimes \Phi_i^{\sigma^{-1} (3)}$ for $\Phi = \Phi^{123} = \sum_i \Phi_i^1 \otimes \Phi_i^2 \otimes \Phi_i^3$ and $\sigma\in S_3$ (see for instance \cite{QuasiHopf}). We can easily see that for the coassociative case, this simplifies to the ordinary Yang-Baxter equation.\par
For the $R$-matrix we introduced in section~\ref{CoproductJuan}, one can indeed prove that the generalised Yang-Baxter equation holds for the above coassociator. More interestingly, even though we are dealing with a non-coassociative case, this $R$-matrix also satisfies the regular Yang-Baxter equation. This rather curious result is owed to the structure of the coassociator. Although $\Phi$ is not proportional to $\alg{1} \otimes \alg{1} \otimes \alg{1}$, it is diagonal and thus rather simple. The hands-on computation of the generalised Yang-Baxter equation revealed that, evaluated on 3-particle states, each of its components is indeed equal to its regular counterpart times $q^{E/2}$ factors. To be more precise, let us start with a given 3-particle state $|\chi_1 \chi_2 \chi_3 \rangle$ and let us denote  the generalised Yang-Baxter equation
\begin{align*}
\textbf{GYBE}  := R_{12} \Phi^{312} R_{13} {\Phi^{132}}^{-1} R_{23} \Phi^{123} - \Phi^{321} R_{23} {\Phi^{231}}^{-1} R_{13} \Phi^{213} R_{12} \ .
\end{align*}
A lengthy calculation shows that, indeed, we always encounter the property
\begin{align*}
\langle \chi_1' \chi_2' \chi_3' | \exp (-\omega_{12} \otimes \mathcal{B}) \textbf{GYBE} \exp (-\omega_{12} \otimes \mathcal{B}) |\chi_1 \chi_2 \chi_3 \rangle
&=\langle \chi_1' \chi_2' \chi_3' | \textbf{YBE} |\chi_1 \chi_2 \chi_3 \rangle \ ,
\end{align*}
$\forall ~ |\chi_1' \chi_2' \chi_3' \rangle$ and $|\chi_1 \chi_2 \chi_3 \rangle$ 3-particle states, where
\begin{align*}
\textbf{YBE} &:= R_{12} R_{13} R_{23} - R_{23} R_{13} R_{12} \ ,
\end{align*}
denotes the regular Yang-Baxter equation. In particular, this implies that, indeed, for our case the $\textbf{GYBE}$ and $\textbf{YBE}$ are equivalent.

Finally, we can see that the quasi-Hopf structure we have encountered can be reduced to a twist. If a coassociator is induced by a twist $\mathcal{F}$, it can be written as \cite{z2,QuasiHopf}
\begin{equation}
	\Phi= \mathcal{F}_{23} (\mathfrak{1} \otimes \Delta)(\mathcal{F}) (\Delta \otimes \mathfrak{1})(\mathcal{F}^{-1}) \mathcal{F}^{-1}_{12} \ ,
\end{equation}
from which is easy to check that the choice $\mathcal{F}=q^{\mathcal{B} \otimes \alpha - \alpha \otimes \mathcal{B}}$ generates a coassociator of the form
\begin{equation}
\label{coassociator_concrete}
	\Phi=q^{\mathcal{B} \otimes (\Delta \alpha - \mathfrak{1} \otimes \alpha - \alpha \otimes \mathfrak{1}) +(\Delta \alpha - \mathfrak{1} \otimes \alpha - \alpha \otimes \mathfrak{1}) \otimes \mathcal{B}} \ ,
\end{equation}
for any $[\alpha , \mathcal{B}]=0$. We can see that the coassociator~(\ref{prova}) corresponds to setting $\alpha = \frac{E (p)}{8 i}$. Undoing this twist implies stripping the coproduct entirely of all $q^{E/4}$ factors, leaving only the braiding factors that depend on the momentum. We can clearly see in this example how twists dramatically alter the physics, since removing the non-coassociativity lands us in a completely different model. Nevertheless, we can check that the $R$-matrix obtained after undoing this twist $\tilde{R}=\mathcal{F}_{21}^{-1} R \mathcal{F}_{12}$ (which is still $q$-dependent, due to the $q$-deformed algebra representation) fulfils the YBE.

\section{Conclusions}

In this article we have studied the boost operator both in the regular $\alg{su} (1|1)$ algebra and in the $q$-deformed $U_q [\alg{su} (1|1)]$ algebra, related to string theories in $AdS_3\times S^3$ and $(AdS_3\times S^3)_\eta$ backgrounds, respectively, restricting ourselves to massless particles.

Regarding the $\alg{su} (1|1)$ algebra, we have constructed the coproduct of the boost operator for the most general allowed coproduct of the algebra. The possible coproducts we can construct for an algebra are restricted to those that give rise to a consistent $R$-matrix, in the sense that it should be unitary, braided-unitary and fulfil the Yang-Baxter equation. We can divide the possible coproducts into those which have trivial coproduct for the Cartan element (denoted by \emph{bosonically unbraided family}) and those which have a non-trivial coproduct for the Cartan element (denoted by \emph{bosonically braided family}). These two are physically inequivalent, meaning that they are the only two families which allow for a cocommutative energy-coproduct, and they cannot be mapped into one another by parity transformations (being that the reason why we count the two bosonically braided families as one) or rescaling of the generators ({\it i.e.} one-particle transformations, not allowing for Hopf-algebraic two-particle twists which in general change the physics). We have found coproducts for the boost operator in both of these families of coproducts. In previous articles \cite{JoakimAle,Fontanella:2016opq,FontanellaTorrielli} a coproduct was found which, although perfectly legitimate from an algebraic standpoint, {\it annihilate} the $R$-matrix. What we find here is more akin to what \cite{BorStromTorri} studied for massive representations: these coproducts {\it quasi-cocommute} with the correspondent $R$-matrix in their family. In this sense they are a true symmetry of the correspondent $R$-matrix, as opposed to generating a differential equations of the type $\mbox{[$1^{\mbox{st}}$ order diff. operator in $(p_1,p_2)$]} R(p_1,p_2) = 0$. Furthermore, we have found that we are not able to completely fix these coproducts and they present some ambiguities. We have no physical arguments we can use to fix them, as they commute with (the coproduct of) all the elements of the algebra and quasi-commute with the $R$-matrix. However, these ambiguities are intimately related with the $R$-matrix, as they seem to give rise to some kind of evolution equation for it. This evolution equation turns out to be ultimately equivalent, for the case where the comparison is possible, to the differential equation we find for the coproduct which annihilate the R-matrix.

Finally, only the family with unbraided energy-coproduct is directly connected to $AdS_3$ superstrings. The law of addition of multi-particle energies deriving from it is simply $\alg{H}_1+\alg{H}_2$ as expected for magnons. If one considers the braided energy-coproduct as studied in \cite{JoakimAle}, one can curiously identify it with the energy of two phonons on a linear chain undergoing umklapp scattering, as pointed out in previous literature on the $q$-Poincar\'e structure underlying such processes \cite{CeleghiniEtAl}. The braided coproduct in fact has the following property: $\Delta(\alg{H}) = \alg{H}(p_1+p_2) = \alg{H}\big(\Delta(p)\big)$ \cite{CeleghiniEtAl,JoakimAle}.

Regarding the $U_q [\alg{su} (1|1)]$ algebra, we have found that a consistent boost operator can be defined for this algebra. We have constructed its coproduct for one particular choice of the coproduct of the algebra and we have found that the same kind of ambiguities that arises in the non-$q$-deformed algebra also appear here. Despite that, we were not able to find a relation that involves this ambiguity and the $R$-matrix. We have also studied a second particular choice of coproduct for this algebra with less success, as in this case we were not able to construct a coproduct of the boost that quasi-cocommutes with the full $R$-matrix. We do not completely understand why the construction fails here, but we think it might be related to the non-coassociativity of this second coproduct. Although this non-coassociativity breaks the Hopf algebra structure, the coproduct satisfies a more general structure called \emph{quasi-Hopf algebra}. It would be interesting to explore more in depth this structure and its physical implications, which might help us understand why we were not able to find a consistent coproduct of the boost in this case.

One open question related to our findings is how the ambiguities give rise to equations for the $R$-matrix of the form
\begin{equation}
	(\partial_{p_1} -\partial_{p_2}) R + [ \Upsilon , R]=0 \ , \qquad (\partial_{p_1} -\partial_{p_2}) R + [ \beta , R]=0 \ ,
\end{equation}
for the bosonically unbraided and bosonically braided families of coproducts, respectively. We have observed that the operators $\Upsilon - \Upsilon^{op}$ and $\beta - \beta^{op}$ fulfil the conditions for being a connection for the parallel transport of an $R$-matrix enumerated in \cite{FontanellaTorrielli}. We would like to find a first-principle explanation of why these ambiguities behave in such way. This would also help us find a similar relation for the $q$-deformed case, which we were not able to find.

Another interesting direction we would like to explore is the ambiguity created by the coproduct of $\sin p$. We have not been able to fix it either from first principle or from physical considerations, so we wonder if it can be used to impose further constraints on the form of the $R$-matrix.

It would be worth analysing a different representation of the boost. Instead of writing in in terms of the derivative of the momentum, we can write it in terms of the derivative of the energy. Assuming that we can construct the coproduct of the boost in this representation, it might provide us information about the $R$-matrix of the mirror model in the same fashion as this representation give us information about the regular $R$-matrix.

Finally, this whole construction has been done in the massless limit. Due to their origin, we expect that similar ambiguities in the definition of the coproduct of the boost exist for the massive case. Thus, we plan to revisit the analysis performed in \cite{BorStromTorri} from this new perspective.

\section{Acknowledgements}

We thank Bogdan Stefa\'nski for discussions and for raising the question about the relationship between difference form and braiding factors. We thank Riccardo Borsato for very useful discussions. We thank Riccardo Borsato, Rafael Hernández, Ben Hoare, Roberto Ruiz and Bogdan Stefa\'nski, for very useful comments on the manuscript. We thank the anonymous referee for very useful comments. LW is funded by a University of Surrey Doctoral College Studentship Award. This work is supported by the EPSRC-SFI grant EP/S020888/1 {\it Solving Spins and Strings}. 

No data beyond those presented and cited in this work are needed to validate this study.

\appendix

\section{\label{diffform} $R$-matrix and difference form}

An $R$-matrix is of difference form if there exists a variable $\gamma=\gamma (p)$ such that $R(\gamma_1 , \gamma_2 )=R(\gamma_2 - \gamma_1)$, which implies that
\begin{equation}
	\partial_{\gamma_1} R+\partial_{\gamma_2} R=f(p_1) \partial_{p_1} R + f(p_2) \partial_{p_2} R=0 \ ,
\end{equation}
where $f(p)=\frac{\partial p}{\partial \gamma}$. If we take the logarithm of the previous expression and further differentiate with respect to either of the momenta,we get that any given element of the $R$-matrix, $R_{ab}^{cd}$ should satisfy
\begin{equation}
	\partial_{p_1} \log f(p_1)=-\partial_{p_1} \log \left(\frac{\partial_{p_2} R_{ab}^{cd}}{\partial_{p_1} R_{ab}^{cd}} \right) \ , \label{differenceformcondition}
\end{equation}
and similarly for $p_2$. Notice that the left hand side of this expression is independent of $p_2$, implying that the right hand side is also independent of $p_2$ for any given $R_{ab}^{cd}$, if that component of the $R$-matrix is of difference form.

Furthermore, this construction provides us with a necessary condition whether or not a specific $R$-matrix entry can be recast in difference form, and, if so, it tells us what the variable is in which such entry is of difference form. In particular
\begin{equation}
	\gamma (p_1) = \int{\frac{\partial_{p_1} R_{ab}^{cd}}{\partial_{p_2} R_{ab}^{cd}} g(p_2) \, d p_1} \ ,
\end{equation}
where $g(p_2)$ is a multiplicative factor that makes the product $\frac{\partial_{p_1} R}{\partial_{p_2} R} g(p_2)$ independent of $p_2$. Notice that a function with this property will always exists if the right hand side of equation~\ref{differenceformcondition} is independent of $p_2$. If this change of variables is the same for every element of the $R$-matrix, we can write it in difference form.

We have applied this formalism to the $R$-matrices we have studied in section~\ref{Generalbraiding}. All of them satisfy equation~\ref{differenceformcondition} and give rise to a consistent change of variables for the diagonal entries. However, only in the bosonically braided case with $x+y=0$ and the bosonically unbraided cases can this be extended to the non-diagonal terms, and only for a particular value for each of the braidings. Interestingly, the only case which we cannot write as a difference form, the factor appear in a separate and multiplicative way. Thus, all the $R$-matrices we have studied in section~\ref{Generalbraiding} can be written in difference form after a correct redefinition of the fermionic charges.

\subsection{Undeformed bosonically braided}

After following the procedure detailed above, we find that the change of variables $p=4\arccot e^\gamma$ bring the $R$-matrix associate to the bosonically braided family of coproducts with $x+y=0$ to an $R$-matrix of the form
\begin{equation}
	R_{x+y=0}=\left(\begin{matrix}
	1 & 0 & 0 & 0 \\ 0 & -\tanh \left( \frac{\gamma_1 - \gamma_2}{2} \right)  & e^{-\frac{i}{4} (p_1-p_2) x} \sech \left( \frac{\gamma_1 - \gamma_2}{2} \right) & 0 \\ 0 & e^{\frac{i}{4} (p_1-p_2) x} \sech \left( \frac{\gamma_1 - \gamma_2}{2} \right) & \tanh \left( \frac{\gamma_1 - \gamma_2}{2} \right) & 0 \\ 0 & 0 & 0 & -1
\end{matrix}	  \right) \ .
\end{equation}
Regarding the subfamilies $x+y=\pm 2$, the $R$-matrices cannot be written in difference form in this case, but the change of variables $p=2 \arccot e^\gamma$ make the diagonal entries of difference form
\begin{equation}
	R_{x+y=\pm 2}=\left(\begin{matrix}
	1 & 0 & 0 & 0 \\ 0 & -\tanh \left( \frac{\gamma_1 - \gamma_2}{2} \right)  & \frac{e^{-\frac{i}{4} (p_1-p_2) (x\mp 1)}}{ \cosh \left( \frac{\gamma_1 - \gamma_2}{2} \right) } \left( \frac{\cos \frac{p_1}{2}}{\cos \frac{p_2}{2}} \right)^{\mp 1/2}& 0 \\ 0 & \frac{e^{\frac{i}{4} (p_1-p_2) (x\mp 1)}}{ \cosh \left( \frac{\gamma_1 - \gamma_2}{2} \right) } \left( \frac{\cos \frac{p_1}{2}}{\cos \frac{p_2}{2}} \right)^{\pm 1/2}  & \tanh \left( \frac{\gamma_1 - \gamma_2}{2} \right) & 0 \\ 0 & 0 & 0 & -1
\end{matrix}	  \right) \ .
\end{equation}

\subsection{Undeformed bosonically unbraided}

For the case of bosonically unbraided coproducts with $a=-b=c=-d$, the transformation we are looking for is $p=2\arcsin (e^\gamma )$. After implementing it, the $R$-matrix takes the form
\begin{equation}
	R=\left(\begin{matrix}
	1 & 0 & 0 & 0 \\ 0 & -\tanh \left( \frac{\gamma_1 - \gamma_2}{2} \right)  & e^{-\frac{i}{4} a (p_1-p_2)} \sech \left( \frac{\gamma_1 - \gamma_2}{2} \right) & 0 \\ 0 & e^{\frac{i}{4} a (p_1-p_2)}\sech \left( \frac{\gamma_1 - \gamma_2}{2} \right) & \tanh \left( \frac{\gamma_1 - \gamma_2}{2} \right) & 0 \\ 0 & 0 & 0 & -1
\end{matrix}	  \right) \ .
\end{equation}

For the case of bosonically unbraided coproducts with $a=-b=c\pm 2=-d\pm 2$, the transformation $p=4\arccot (e^\gamma )$ allow us to write the $R$-matrix as
\begin{equation}
	R_{a=c\pm 2}=\left(\begin{matrix}
	1 & 0 & 0 & 0 \\ 0 & -\tanh \left( \frac{\gamma_1 - \gamma_2}{2} \right)  & e^{\frac{i}{4} (1\mp a) (p_1-p_2)} \sech \left( \frac{\gamma_1 - \gamma_2}{2} \right) & 0 \\ 0 & e^{-\frac{i}{4} (1\mp a) (p_1-p_2) } \sech \left( \frac{\gamma_1 - \gamma_2}{2} \right) & \tanh \left( \frac{\gamma_1 - \gamma_2}{2} \right) & 0 \\ 0 & 0 & 0 & -1
\end{matrix}	  \right) \ .
\end{equation}

\section{\label{etakinematics} Action of the boost generator in the $U_q [\alg{sl}(1|1)]$ algebra}

In this appendix we justify the form of the algebra and the ($q$-deformed) modified Poincaré structure we used in section~\ref{Deformedbraiding} to describe excitations in $(AdS_3 \times S^3)_\eta$. First of all, we choose the commutation relation between the supercharges to be of the standard form of a $q$-deformed algebra
\begin{equation}
\label{41}
\{\alg{Q}_{\smallR}, \alg{S}_{\smallR}\} \ = \ [\alg{H}_{\smallR}]_q, \quad \{\alg{Q}_{\smallL} \ , \alg{S}_{\smallL}\} \ = \ [\alg{H}_{\smallL}]_q \ .
\end{equation}

In order to compute the action of the boost operator on the momentum and the energy we should start by expressing them in terms of an uniformising rapidity. The dispersion relation for excitation in this deformed setting is given by
\begin{equation}
\epsilon^2 [E/2]^2_{q}- h^2\sin^2\frac{p}{2}=\left[m/2\right]_q^2 \ , \qquad \epsilon^2= 1-\frac{h^2}{4}(q-q^{-1})^2 \ , \label{defdispersion}
\end{equation}
where $h$, $m$ and $q$, respectively, are the coupling constant, the mass and deformation parameter, and
\begin{equation}
\left[ x \right]_q \equiv \frac{q^x - q^{-x}}{q-q^{-1}} \ . \label{q-analog}
\end{equation}
We can get rid of the mass parameter if we perform the substitutions $q\rightarrow q^m$, $h^2\rightarrow h^2 [m]_q$ and $E\rightarrow E/m$. This allow us to borrow the construction of the uniformising torus for the $(AdS_5 \times S^5)_\eta$ space (see, for example, \cite{RapidityTorus}), and write the momentum and the energy in terms of just one variable $\z$ as
\bea
q^H&=& \frac{\cs (\z, \kappa )+i\dn (\z_0, \kappa )}{\cs(\z, \kappa )-i\dn (\z_0, \kappa )}=e^{i({\rm am} (\z+\z_0)+{\rm am}(\z-\z_0))}\ , \\
e^{ip}&=&\frac{\cs (\z_0, \kappa )+i\dn (\z, \kappa )}{\cs (\z_0, \kappa )-i\dn (\z, \kappa )}=e^{i({\rm am} (\z+\z_0)-{\rm am}(\z-\z_0))} \ ,
\eea
where parameters $\z_0$ and $\kappa$ are related to the coupling constant and the deformation parameter as
\begin{equation}
	h=-\frac{i\k}{2 \dn (\z_0, \kappa )}\sqrt{1-\k^2 {\rm sn (\z, \kappa )}^4_0} \ , \qquad q=e^{i\, {\rm am}(2\z_0)}=\frac{\cs (\z_0, \kappa )+i\dn (\z_0, \kappa )}{\cs (\z_0, \kappa )-i\dn (\z_0, \kappa )}\ .
\end{equation}
In the previous expressions sn, cn and dn are the three fundamental Jacobi elliptic functions, am is the elliptic amplitude, and cs$(z,\kappa)=\frac{\text{cn}(z,\kappa)}{\text{sn}(z,\kappa)}$.

From these equations we get that
\begin{align}
	p &= {\rm am} (\z+\z_0 , \kappa ) + {\rm am} (\z+\z_0 , \kappa ) \ , & E &= \frac{{\rm am} (\z+\z_0 , \kappa ) - {\rm am} (\z+\z_0 , \kappa )}{{\rm am} (2 \zeta_0)} \ , \\
	\partial_\z p &= {\rm dn} (\z+\z_0 , \kappa ) + {\rm dn} (\z+\z_0 , \kappa ) \ , & [E]_q &= \frac{{\rm dn} (\z, \kappa )}{{\rm dn (\z_0, \kappa )}} \frac{{\rm cs}^2 (\z_0, \kappa ) + {\rm dn}^2 (\z_0, \kappa )}{{\rm cs}^2 (\z_0, \kappa )+{\rm dn}^2 (\z, \kappa )} \ .
\end{align}
where we have used that $\partial_\z {\rm am} (\z,\kappa )={\rm dn} (\z,\kappa )$. To relate $[\partial_\z , p]=\partial_\z p$ with $[E]_q$ we need the following identities involving elliptic functions
\begin{align*}
	&{\rm dn} (\z+\z_0) = \frac{ {\rm dn} (\z, \kappa ) \ {\rm dn} (\z_0, \kappa ) -\kappa^2 {\rm sn} (\z, \kappa ) \ {\rm dn} (\z, \kappa ) \ {\rm sn} (\z_0, \kappa ) \ {\rm dn} (\z_0, \kappa )}{1-\kappa^2 \ {\rm sn}^2 (\z, \kappa ) \ {\rm sn}^2 (\z_0, \kappa ) } \ , \\
	&1-{\rm dn}^2 (\z, \kappa )= \kappa^2 \left[ 1- {\rm cn}^2  (\z, \kappa )\right] =\kappa^2 {\rm sn}^2 (\z, \kappa ) \ , \\
	&1-\kappa^2 {\rm sn}^2 (\z, \kappa ) {\rm sn}_0^2={\rm sn}^2 (\z_0, \kappa ) \left[ {\rm cs}^2 (\z_0, \kappa ) + {\rm dn}^2 (\z, \kappa ) \right] = {\rm sn}^2 (\z, \kappa ) \left[ {\rm cs}^2 (\z, \kappa ) + {\rm dn}^2 (\z_0, \kappa ) \right] \ ,
\end{align*}
and, after some algebra, we get
\begin{align}
	[\partial_\z , p] &= \frac{ 2 {\rm dn} (\z, \kappa ) \ {\rm dn} (\z_0, \kappa ) }{1-\kappa^2 \ {\rm sn}^2 (\z, \kappa ) \ {\rm sn}^2 (\z_0, \kappa ) } =\frac{ 2 {\rm dn} (\z, \kappa ) \ {\rm dn} (\z_0, \kappa ) }{{\rm cn}^2 (\z_0, \kappa )+ {\rm dn}^2 (\z, \kappa ) \ {\rm sn}^2 (\z_0, \kappa ) } \notag \\
	&= \frac{ 2 [E]_q }{{\rm cd}^2 (\z_0, \kappa )+ {\rm sn}^2 (\z_0, \kappa ) }= \frac{-\kappa^2 [E]_q}{2 h^2} \ ,
\end{align}
where we have used that $h^2=-\frac{\kappa^2}{4 \ {\rm ds}^2 (\z_0, \kappa )}  [{\rm cs}^2 (\z_0, \kappa )+{\rm dn}^2 (\z_0, \kappa )]$.

We can circumvent the explicit computation of $[\partial_\z , E]$ if we instead use the dispersion relation (\ref{defdispersion}) to relate it with $[\partial_\z , p]$. In particular we get
\begin{equation}
	2\epsilon^2 [E/2]_{q} \partial_\z [E/2]_{q}= h^2\left( 2 \sin \frac{p}{2} \cos \frac{p}{2}\right) \frac{\partial_\z p}{2} \Longrightarrow [\partial_\z , [E/2]_q]=-\frac{[E]_q}{[E/2]_q} \frac{\kappa^2\sin p}{8 \epsilon^2} \ .
\end{equation}
From the definition of the $q$-analog (\ref{q-analog}) we can prove that
\begin{align}
	\partial_\z [E/2]_q &= \frac{q^{E/2}+q^{-E/2}}{q-q^{-1}} \frac{\log q}{2}  \partial_z E \ , & \frac{[E]_q}{[E/2]_q}=q^{E/2}+q^{-E/2} \ ,
\end{align}
therefore
\begin{equation}
	[\partial_\z , E]=-\frac{q-q^{-1}}{\log q} \frac{\kappa^2}{4 \epsilon^2} \sin p \ .
\end{equation}

For simplicity, instead of writing the factors every time, we will write these two commutation relations as
\begin{equation}
	[\alg{J} , p]= \alpha [E]_q \ , [\alg{J} , E]=\frac{q-q^{-1}}{\log q} \beta \sin p \ .
\end{equation} 
where we have defined the boost operator in terms of the derivative of the uniformising variable as $\alg{J}=\frac{i}{2} \partial_\z$. Furthermore, as we will be working in the representation where the left and right algebras act in equivalent ways, we can identify $2\alg{H}=E$.

Finally, we can compute the commutation relations between $\alg{J}$ and the supercharges via the Jacobi identity
\begin{displaymath}
 	\left. \begin{matrix}
	[\alg{J},\alg{Q}]=\phi_Q (p) \alg{Q} \\
	[\alg{J},\alg{S}]=\phi_S (p) \alg{S}
	\end{matrix} \right\} \Longrightarrow [\alg{J},[E/2]_q]=[\alg{J}, \{\alg{Q},\alg{S}\}]=\{[\alg{J},\alg{Q}],\alg{S}\}+\{[\alg{J},\alg{S}],\alg{Q}\}=\left[\phi_Q (p) + \phi_S (p) \right] [E/2]_q \ ,
\end{displaymath}
thus, choosing the most symmetric setting and making the two functions equal, we can write
\begin{equation}
	\phi_Q (p)=\phi_S (p)=\frac{[\alg{J},[E/2]_q]}{2 [E/2]_q}= \frac{\beta [E]_q}{4[E/2]_q^2} \sin p \ .
\end{equation}

This algebra has been derived independently of the mass parameter $m$ thank to the previous redefinition, thus we can extend it from $m=1$ to any value of the mass and, in particular, to the massless case.



\end{document}